\documentstyle[psfig,times]{mn}

\def\etal{et~al.}
\def\spose#1{\hbox to 0pt{#1\hss}}
\def\lta{\mathrel{\spose{\lower 3pt\hbox{$\mathchar"218$}}
     \raise 2.0pt\hbox{$\mathchar"13C$}}}
\def\gta{\mathrel{\spose{\lower 3pt\hbox{$\mathchar"218$}}
     \raise 2.0pt\hbox{$\mathchar"13E$}}}

\title[The host galaxies of radio--loud AGN]{The host galaxies of
radio--loud AGN: mass dependencies, gas cooling and AGN feedback}

\author[P.~N.~Best \etal]{P.~N.~Best,$^1$\thanks{Email:
pnb@roe.ac.uk} G. Kauffmann,$^2$ T. M. Heckman,$^3$ J.
Brinchmann$^4$, S. Charlot$^{5,2}$,\vspace*{0.5mm}\\
\LARGE \v{Z}. Ivezi{\'c}$^6$, S. D. M. White$^2$\\
$^1$ Institute for Astronomy, Royal Observatory Edinburgh, Blackford Hill,
Edinburgh EH9 3HJ, UK\\
$^2$ Max-Planck-Institut f{\"u}r Astrophysik, Karl-Schwarzschild-Str. 1,
Postfach 1317, D-85741 Garching, Germany\\
$^3$ Department of Physics \& Astronomy, The Johns Hopkins University,
Baltimore, MD 21218, USA\\
$^4$ Centro de Astrof{\'\i}sica da Universidade do Porto, Rua das Estrelas,
4150-762 Porto, Portugal\\
$^5$ Institut d'Astrophysique de Paris, CNRS, 98 bis boulevard Arago,
F-75014 Paris, France\\
$^6$ Princeton University Observatory, Peyton Hall, Princeton, NJ
08544-1001, USA}

\pagerange{\pageref{firstpage}--\pageref{lastpage}}
\pubyear{2004}
\begin{document}
\label{firstpage}

\maketitle

\begin{abstract}
\noindent The properties of the host galaxies of a well--defined sample of
2215 radio--loud AGN with redshifts $0.03 < z < 0.3$, defined from the
Sloan Digital Sky Survey (SDSS), are investigated. These are predominantly
low radio luminosity sources, with 1.4 GHz luminosities in the range
$10^{23}$ to $10^{25}$ W\,Hz$^{-1}$. The fraction of galaxies that host
radio--loud AGN with $L_{\rm 1.4GHz} > 10^{23}$W\,Hz$^{-1}$ is a strong
function of stellar mass, rising from nearly zero below a stellar mass of
$10^{10} M_{\odot}$ to more than 30\% at stellar masses of $5 \times
10^{11} M_{\odot}$. In contrast to the integrated [OIII] luminosity
density from emission line AGN, which is mainly produced by black holes
with masses below $10^8 M_{\odot}$, the integrated radio luminosity
density comes from the most massive black holes in the Universe.  The
integral radio luminosity function is derived in six ranges of stellar and
black hole mass. Its shape is very similar in all of these ranges and can
be well fitted by a broken power-law. Its normalisation varies strongly
with mass, as $M_*^{2.5}$ or $M_{\rm BH}^{1.6}$; this scaling only begins
to break down when the predicted radio--loud fraction exceeds 20--30\%.
There is no correlation between radio and emission line luminosities for
the radio-loud AGN in the sample and the probability that a galaxy of
given mass is radio-loud is independent of whether it is optically
classified as an AGN.  The host galaxies of the radio--loud AGN have
properties similar to those of ordinary galaxies of the same mass, with a
tendency for radio--loud AGN to be found in larger galaxies and in richer
environments.  The host galaxies of radio--loud AGN with emission lines
match those of their radio--quiet counterparts.

All of these findings support the conclusion that the optical AGN and low
radio luminosity AGN phenomena are independent and are triggered by
different physical mechanisms. Intriguingly, the dependence on black hole
mass of the radio--loud AGN fraction mirrors that of the rate at which gas
cools from the hot atmospheres of elliptical galaxies. It is speculated
that gas cooling provides a natural explanation for the origin of the
radio--loud AGN activity, and it is argued that AGN heating could
plausibly balance the cooling of the gas over time.
\end{abstract}


\section{Introduction}

In recent years it has become apparent that active galactic nuclei (AGN)
are not only interesting objects to study in their own right, but may also
play an important role in the process of galaxy formation and evolution.
One major development has been the detection of a central (generally
dormant) massive black hole at the heart of essentially all nearby
galaxies. Further, the mass of these black holes is tightly correlated
with that of the stellar bulge component of their host galaxies, as
estimated from the velocity dispersion \cite{fer00,geb00}, the absolute
magnitude of the bulge \cite{kor95,mag98a} or using virial methods
\cite{mar03}. This correlation indicates that the build-up of a galaxy and
that of its central black hole are fundamentally linked. Theoretical
interpretations of this favour models in which the host galaxy and its
central black hole grow simultaneously up until the point when either all
of the gas has been consumed (e.g. Archibald \etal\ 2002)\nocite{arc02a}
or the central AGN has reached sufficient luminosity that its winds are
able to drive the remaining gas out of the galaxy, thereby terminating
both star formation and black hole growth
\cite{sil98,fab99,kin03}.

Highly luminous AGN were considerably more common in the early Universe
than they are today; their co-moving space density has fallen by about a
factor of 1000 since redshifts 2--3 (e.g.\ Hartwick \& Schade
1990)\nocite{har90}. Nevertheless, much can still be learned about AGN
through studies of their host galaxies in the nearby Universe. In
particular, the new large galaxy redshift surveys, especially the 2-degree
Field Galaxy Redshift Survey (2dFGRS; Colless \etal\ 2001)\nocite{col01}
and the Sloan Digital Sky Survey (SDSS; York \etal\ 2000; Stoughton \etal\
2002)\nocite{yor00,sto02} allow the definition of sufficiently large
samples of nearby AGN to enable comprehensive statistical analyses of
their host galaxy properties to be carried out.

Kauffmann \etal\ (2003a; hereafter K03)\nocite{kau03c} constructed a
sample of 22,623 optically--selected narrow--line AGN from the first data
release (DR1) of the SDSS main spectroscopic catalogue. They investigated
in detail various properties of the host galaxies of these AGN, such as
their stellar masses, sizes, surface mass densities, mean stellar ages,
and past star formation histories (see Kauffmann \etal\ 2003b,c for a
description of how these properties were derived).\nocite{kau03a,kau03b}
They found that optically--selected AGN reside almost exclusively in
massive galaxies, with the fraction of AGN decreasing rapidly below a
stellar mass of $10^{10} M_{\odot}$. The structural properties of the AGN,
such as their sizes and stellar mass densities, were found to be
statistically indistinguishable from those of normal early--type galaxies
of the same mass. For low emission--line luminosity AGN, which are
predominately Low--Ionisation Nuclear Emission--line Region (LINER) type
galaxies, this was also true of their stellar populations. The higher
emission line luminosity AGN ($L_{\rm [OIII] 5007} \gta 10^7 L_{\odot}$;
these are predominantly Seyferts), however, were found to have lower
4000\AA\ break strengths [hereafter $D_n(4000)$], implying on-going star
formation, and to exhibit evidence of galaxy--wide bursts of star
formation in the recent past. Indeed, these high emission line luminosity
AGN occupy a region of the stellar mass ($M_*$) versus $D_n(4000)$ plane
which is poorly populated by ordinary galaxies.  K03\nocite{kau03c}
concluded that emission line AGN and starburst activity are tightly related
phenomena (cf. Sanders \etal\ 1988).\nocite{san88} They argued that this
may have a straightforward interpretation since powerful AGN require two
key ingredients: a massive black hole and a supply of gas to fuel
it. Massive black holes are only found in galaxies with massive bulges,
and where there is an ample supply of gas it is not unreasonable to
suppose that there will also be on-going star formation.

One issue that needs to be considered, however, is that in the K03
\nocite{kau03c} study the AGN were optically selected by their emission
line properties, using the [OIII]~5007 / H$\beta$ versus [NII]~6583 /
H$\alpha$ emission line ratio diagnostic diagram \cite{bal81} to separate
out the AGN from those galaxies where the emission lines are associated
with star formation. As such, only galaxies in which all four of these
emission lines are detected can be classified as AGN, meaning that the
sample is naturally biased in favour of AGN with bright emission
lines. Given the importance of understanding the physical nature of the
AGN phenomenon, it is clearly necessary to test the robustness of the
emission--line selection technique and determine whether emission--line
selection influences conclusions about the nature of AGN host galaxies,
the growth of black holes and their associated energetic output.

Radio--loud AGN are generally hosted by giant elliptical galaxies
(e.g. Matthews, Morgan, \& Schmidt 1964)\nocite{mat64}.  The most powerful
radio sources (the `classical doubles', known as Fanaroff \& Riley
[1974]\nocite{fan74} Class II sources, or FR\,IIs) usually have strong
line emission, which is roughly proportional to their radio luminosity
(e.g. Rawlings \etal\ 1989)\nocite{raw89}. However, it has long been known
that a subset of FR\,IIs have only very weak or no emission lines
\cite{hin79}, as do most of the lower radio luminosity FR\,Is (`edge--darkened'
radio sources, whose jets decelerate significantly and flare out in the
inner kiloparsec probably due to entrainment of material). Sadler \etal\
\shortcite{sad02} defined a sample of radio--loud AGN within the 2dFGRS
and found that of order half of them (the exact fraction depends upon the
quality of the spectra) have absorption line spectra similar to those of
inactive giant elliptical galaxies, and would be mostly missed by
emission--line selection. There remains much debate as to whether
weak--lined FR\,Is and the possibly related weak--emission lined FR\,IIs
have a different accretion mechanism to their more powerful counterparts
(see Cao \& Rawlings 2004 and references therein)\nocite{cao04}. The
interstellar medium of these galaxies is hot and diffuse, in contrast to
the cold dense interstellar medium of the strong emission line AGN, which
may imply a difference in the way in which the black hole is fueled and
ionising radiation is produced.

Despite their low emission line luminosities, FR\,I radio sources can
still provide a very important energetic output into their environment,
through the expansion of their radio lobes. This process can lead to very
efficient coupling of the jet kinetic energy output of the radio sources
with the interstellar or intergalactic medium.  This has been observed
through the detection of buoyant gas bubbles (e.g. B{\"o}hringer \etal\
1993)\nocite{boh93} and energy--dispersing sound waves in clusters of
galaxies \cite{fab03}. Fabian \etal\ argued that the heating effects of
the weak shocks and sound waves associated with the radio source AGN
activity may work to balance the effects of gas cooling within the central
regions of clusters, and thus explain why signatures of cooler gas are not
observed.

The important new insights into the nature of the emission--line AGN
population provided by the superb statistics in the SDSS provide
encouragement that similarly important insights will emerge from an
analysis of the SDSS population of radio--loud AGN. The goal of the
current paper is to investigate this, by studying the sample of 2215
radio--loud AGN that were defined in the accompanying paper (Best \etal\
2005; hereafter Paper I)\nocite{bes05a}. The radio--loud AGN host galaxies
are compared with those of both optically selected AGN and inactive
galaxies\footnote{In this paper, `inactive' galaxies are taken to mean
those with no observable AGN activity at optical or radio wavelengths;
`optically inactive' galaxies are those with no optical AGN activity, but
which possibly do show radio emission.}, in order to investigate the
nature of AGN in the nearby Universe, to cross--compare different AGN
selection methods, and to investigate the origin of radio loudness.

The layout of the paper is as follows. In Section~\ref{sampdef} a brief
description is provided of the three surveys from which the radio--loud
AGN is defined; the SDSS, the National Radio Astronomy Observatory (NRAO)
Very Large Array (VLA) Sky Survey (NVSS; Condon \etal\ 1998)\nocite{con98}
and the Faint Images of the Radio Sky at Twenty centimetres (FIRST) survey
\cite{bec95}. The reader is referred to Paper I for a full description of
the sample definition process.  Section~\ref{massdep} then studies how the
radio and optical AGN fractions depend upon the mass of the host
galaxy. The inter--relation of radio and optical activity is investigated
in Section~\ref{radoptindep}.  Section~\ref{envdep} discusses briefly how
galaxy environments differ between radio--loud AGN with and without
emission lines. Section~\ref{hostprops} then compares the properties of
the host galaxies of the radio--loud AGN with those of inactive
galaxies. The results are discussed and conclusions drawn in
Section~\ref{discuss}; in particular, the mass dependence of the AGN
activity is compared to the gas cooling rate and to the balance between
cooling and AGN heating. Throughout the paper, the values adopted for the
cosmological parameters are $\Omega_m = 0.3$, $\Omega_{\Lambda} = 0.7$,
and $H_0 = 70$\,km\,s$^{-1}$Mpc$^{-1}$.

\section{The radio--loud AGN sample}
\label{sampdef}

The basis sample for this study is a sample of galaxies described by
Brinchmann \etal\ \shortcite{bri04b}, which was drawn from the `main
galaxy catalogue' of the second data release (DR2) of the Sloan Digital
Sky Survey (York \etal\ 2000; Stoughton \etal\ 2002, and references
therein)\nocite{yor00,sto02}.  This consists of about 212,000 galaxies
with magnitudes $14.5 < r < 17.77$, for which spectroscopic redshifts have
been determined. As described by Brinchmann \etal\ \shortcite{bri04b},
many properties of these galaxies have been parameterised from the imaging
and spectroscopic data, with catalogues of parameters available on the web
(see http://www.mpa-garching.mpg.de/SDSS/).  Measured parameters include:
galaxy sizes, concentration indices, 4000\AA\ break strengths, H$\delta$
absorption measurements \cite{kau03a,kau03b}; accurate emission line
fluxes, after subtraction of the modelled stellar continuum
(K03,\nocite{kau03c} Tremonti \etal\, in preparation); parameters
measuring optical AGN activity, such as emission line ratios, and galaxy
velocity dispersions (hence black hole mass estimates; K03, Heckman \etal\
2004).\nocite{kau03c,hec04} Derived parameters include total stellar
masses, stellar surface mass densities, mass-to-light ratios, dust
attenuation measurements, star formation rates \cite{bri04b}, and
gas--phase metallicities \cite{tre04}. These parameters have been adopted
for the analyses of this paper: the reader is referred to the papers
referenced above for detailed information about the methods used to derive
them.

The NVSS \cite{con98} and FIRST \cite{bec95} radio surveys have both been
carried out in recent years using the VLA radio synthesis telescope at a
frequency of 1.4\,GHz, but at differing angular resolutions. The NVSS
covers the entirety of the sky north of $-40^{\circ}$ declination, at an
angular resolution of 45 arcseconds, down to a limiting point source flux
density of about 2.5\,mJy. It is therefore sensitive to extended radio
emission, but its angular resolution is a little poor for reliable
cross--identification of the sources with their optical counterparts. The
FIRST observations cover a sky area designed to largely overlap with that
of the SDSS, down to a limiting flux density of about 1\,mJy for point
sources. These have a much higher angular resolution of $\sim 5$ arcsec,
allowing reliable cross-comparison of point sources, but meaning that
extended sources are often resolved into multiple components, and some (or
even all) of the flux density may be resolved out.

The FIRST and NVSS radio surveys are highly complementary for identifying
radio sources associated with nearby galaxies. In Paper I an algorithm was
developed to identify radio sources within the SDSS spectroscopic sample
using a hybrid method, which optimised the advantages of the two
individual surveys whilst avoiding most of the completeness and
reliability problems associated with the use of either survey
individually. In this way it was possible to derive a sample of 2712
radio--loud galaxies to a 1.4\,GHz flux density limit of 5\,mJy, with the
sample having a completeness of approximately 95\% and a reliability of
98.9\%. The location of these radio--loud galaxies in the plane of
4000\AA\ break strength versus radio luminosity per unit stellar mass was
then used to separate this sample into the radio--loud AGN (2215 galaxies)
and the star--forming galaxies (497).

In the current paper, many analyses of host galaxy properties are
restricted to the subset of 420 radio--loud AGN from that sample with
redshifts $z \le 0.1$. This is for two reasons. First, for the sample
flux--density limit of 5\,mJy, $z=0.1$ corresponds to a radio luminosity
limit of about $10^{23}$W\,Hz$^{-1}$ which represents the majority of the
radio--loud AGN population. Second, at higher redshifts the fraction of
galaxies with detectable emission lines declines sharply, both because of
the increased distance and because the 3-arcsec diameter fibre aperture
includes a larger background of starlight from the host galaxy, so there
is declining sensitivity to weak nuclear emission lines (only [OIII]
luminosities brighter than $\sim 10^{5.8} L_{\odot}$ can be detected at $z
=0.1$). For analyses in which emission line measurements are not required
and it is straightforward to account for the redshift dependence of the
radio luminosity limit, the entire sample of radio--loud AGN is retained
in order to provide improved statistics.

A final issue worth noting is that, as discussed in Paper 1 and by K03,
the use of the SDSS main galaxy sample as the basis sample means that
objects classified as `quasars' by the automated SDSS classification
pipeline (Schlegel \etal\, in preparation) are excluded, even if they lie
in the redshift range under study. The exclusion of these objects is
necessary since the presence of direct light from the active nucleus
prohibits the host galaxy parameters discussed above from being accurately
determined. In orientation--based unification schemes for AGN
(e.g. Antonucci 1993 and references therein)\nocite{ant93}, however, such
AGN (Type-I's) are the same as the (Type-II) AGN under study, but simply
viewed at a different angle to the line of sight. If such unification
schemes are correct then the only effect of restricting study to Type-II
objects is that all AGN fractions derived should be scaled upwards by a
small factor. The size of this scale factor depends upon the relative
numbers of AGN included and excluded from the analysis.

For the low--luminosity AGN under study here, Type-II AGN greatly
outnumber Type-I AGN (although this changes at higher luminosities,
suggesting a luminosity--dependent torus opening angle, e.g. Hao \etal\
2005, Simpson 2005)\nocite{hao05,sim05}.  Further, at these low
luminosities many Type-I AGN are pipeline classified as `galaxies' rather
than `quasars' anyway, since the host galaxy greatly outshines the weak
AGN. These objects therefore remain included in our sample (cf. K03, who
estimate that 8\% of AGN in the main galaxy sample have broad emission
lines, and therefore are strictly Type-Is); only those classified as
quasars are excluded. In the redshift range $0.03 \le z \le 0.10$, in
which most of the analyses of this paper are carried out, a search of the
SDSS DR2 database reveals only 393 objects classified as quasars. This
compares to 16661 objects classified as emission--line AGN by K03 in the
same redshift range. Even allowing for some Type-I AGN being excluded from
spectroscopic study, e.g. because the nuclear emission brightens them out
of the target magnitude range, only a few percent of all AGN are excluded
by the removal of the `quasar' objects. The correlations discovered in
this paper are far too strong to be affected by any such changes, and
therefore the exclusion of some Type-I AGN has no effect upon the results
of this paper.

\section{Mass dependence of AGN activity}
\label{massdep}

The SDSS data allow estimates to be made of both the stellar mass and the
black hole mass of the galaxies. The stellar mass is derived from the
extinction--corrected optical luminosity by using a mass--to--light ratio
that is estimated using the 4000\AA\ break strength and the H$\delta$
absorption index (Kauffmann \etal\ 2003b). Black hole mass ($M_{\rm BH}$)
is a more fundamental parameter for AGN since it probes the central engine
directly. The black hole mass can be estimated using the velocity
dispersion ($\sigma_*$) of the galaxy and the relation between velocity
dispersion and black hole mass given in Tremaine \etal\ \shortcite{tre02}:
log$(M_{\rm BH} / M_{\odot}) = 8.13 + 4.02
\rm{log}(\sigma_*/200$km\,s$^{-1}$)\footnote{This relation is used for
consistency with the emission--line AGN studies. Adoption of a different
black hole mass versus velocity dispersion relation, such as that of
Ferrarese \& Ford \shortcite{fer04}, simply requires a straightforward
re-scaling of the black hole masses.}.  In general, however, black hole
masses are more poorly determined than stellar masses. There are a number
of reasons for this: (i) the velocity dispersion measurements have bigger
errors than the magnitude measurements; (ii) for disk--dominated galaxies,
the velocity dispersion measured within the fibre does not provide an
accurate measure of black hole mass; (iii) perhaps most importantly, there
is significant scatter in the black hole mass versus velocity dispersion
relation. This means that for individual objects, the black hole mass
estimate may be wrong by a factor of a few. In this paper, analyses are
therefore carried out using either stellar mass or black hole mass or
both, as is most appropriate for the property under study.

\begin{figure}
\centerline{
\psfig{file=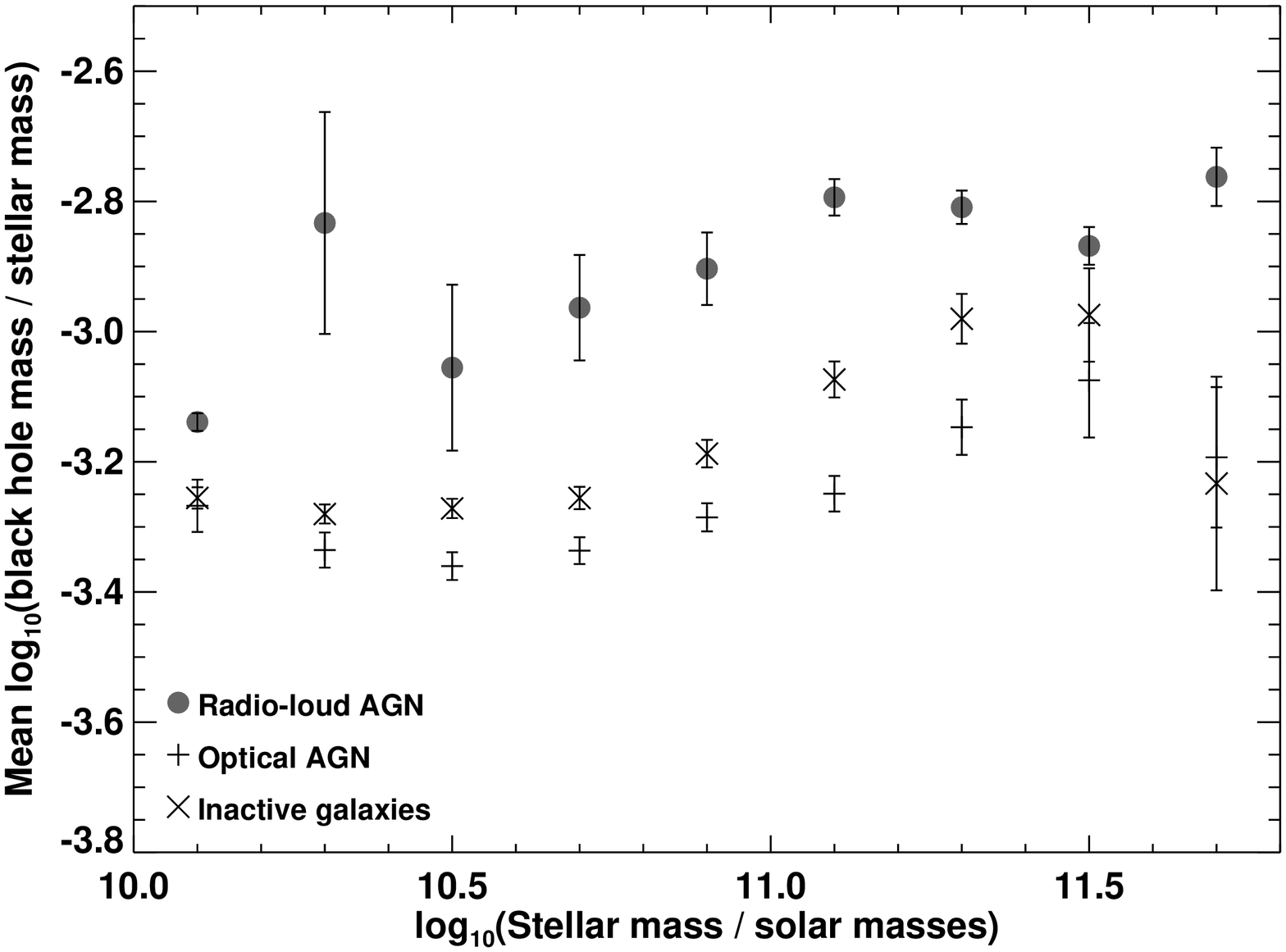,angle=90,width=8.6cm,clip=}
}
\centerline{
\psfig{file=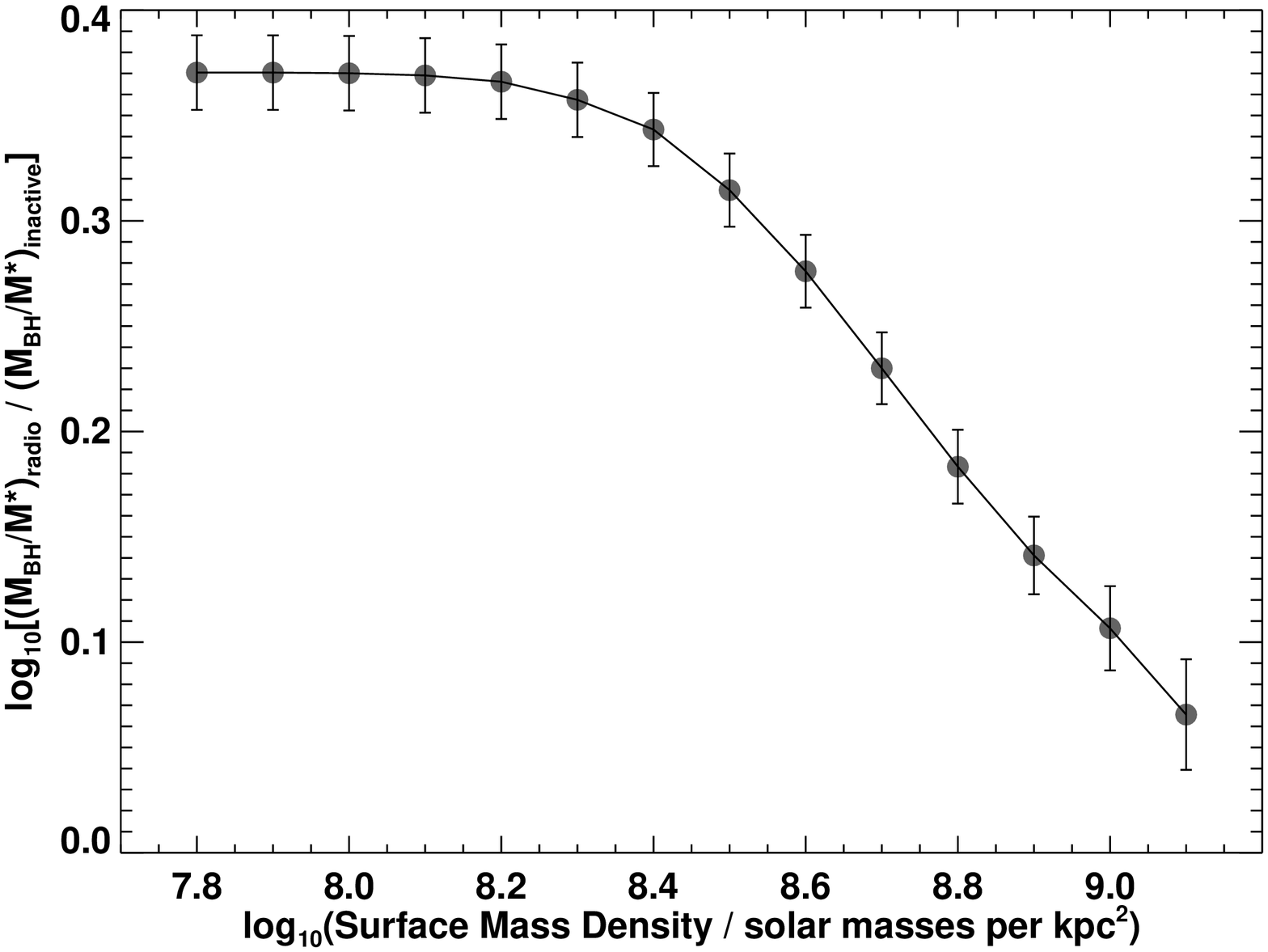,angle=90,width=8.6cm,clip=}
}
\centerline{
\psfig{file=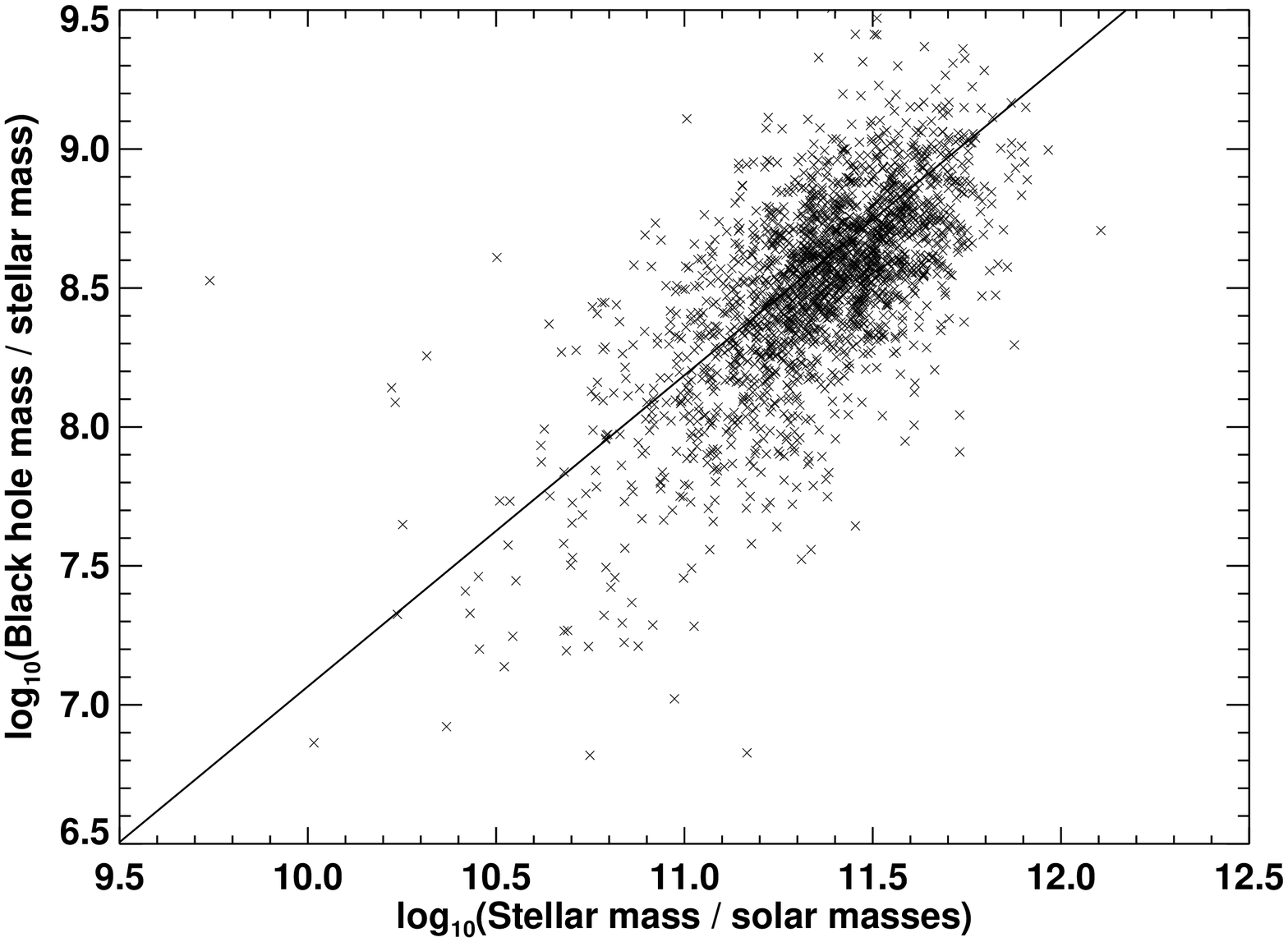,angle=90,width=8.6cm,clip=}
}
\caption{\label{bh_mass} {\it Top:} the mean black hole to stellar mass
ratio as a function of stellar mass, for radio--loud AGN, optical
(emission--line selected) AGN, and inactive galaxies. {\it Middle:} the
mean difference between the black hole to stellar mass ratio of the
radio--loud AGN to that of the inactive galaxies, considering only those
galaxies with surface mass densities above a given value. This
demonstrates that by removing late--type (low surface mass density)
galaxies from the analysis, the difference between the radio--loud AGN and
the inactive galaxies largely disappears. Thus, the higher black hole
masses of the radio--loud AGN arise because these generally lie in
galaxies with larger bulge--to--disk ratios. {\it Bottom:} The black hole
mass versus stellar mass distribution of the radio--loud AGN, compared to
the relation between bulge mass and black hole mass derived by H\"aring \&
Rix \shortcite{har04}: $M_{\rm BH} = 0.0014 M_{\rm bulge} \times (M_{\rm
bulge} / 5\times10^{10} M_{\odot})^{0.12}$. Essentially all of the stellar 
mass of radio--loud AGN is associated with a bulge.}
\end{figure}

Before beginning the analysis, it is worth noting that at a fixed stellar
mass, radio--loud AGN are biased towards larger bulges and larger black
holes. This is demonstrated in Figure~\ref{bh_mass}. The top panel shows
the mean black hole to stellar mass ratio as a function of stellar mass
for radio--loud AGN, emission-line selected AGN and inactive galaxies. The
radio--loud AGN have larger mean black hole masses at a given stellar mass
than inactive galaxies (while for emission--line selected AGN the opposite
is true). The stellar surface mass density ($\mu_*$) of a galaxy can be
used to as a guide to its morphology; as shown by Kauffmann \etal\
\shortcite{kau03a}, the value $\mu_* \sim 3 \times 10^8
M_{\odot}$kpc$^{-2}$ marks the point where the galaxy population undergoes
a transition from disk--dominated to bulge--dominated galaxies.  The
middle panel of Figure~\ref{bh_mass} shows that the difference between the
mean black hole mass to stellar mass ratio of radio--loud AGN and that of
inactive galaxies decreases for galaxies with higher mass surface
densities. This indicates that radio--loud AGN have more massive black
holes (at given stellar mass) because they are in galaxies with larger
bulge--to--disk ratios. Indeed the bottom panel shows the black hole mass
versus stellar mass distribution of the radio--loud AGN; the plotted line
is the relation between bulge mass and black hole mass derived by H\"aring
\& Rix \shortcite{har04}.  Particularly at the higher masses, the
radio--loud AGN are distributed around this relation, indicating that
essentially all of their stellar mass is associated with a bulge.

\subsection{Radio--loud AGN fractions}

It has long been known that radio--loud AGN are preferentially hosted
by massive elliptical galaxies. The SDSS data allow this mass
dependence to be investigated in detail.

\begin{figure*}
\centerline{
\psfig{file=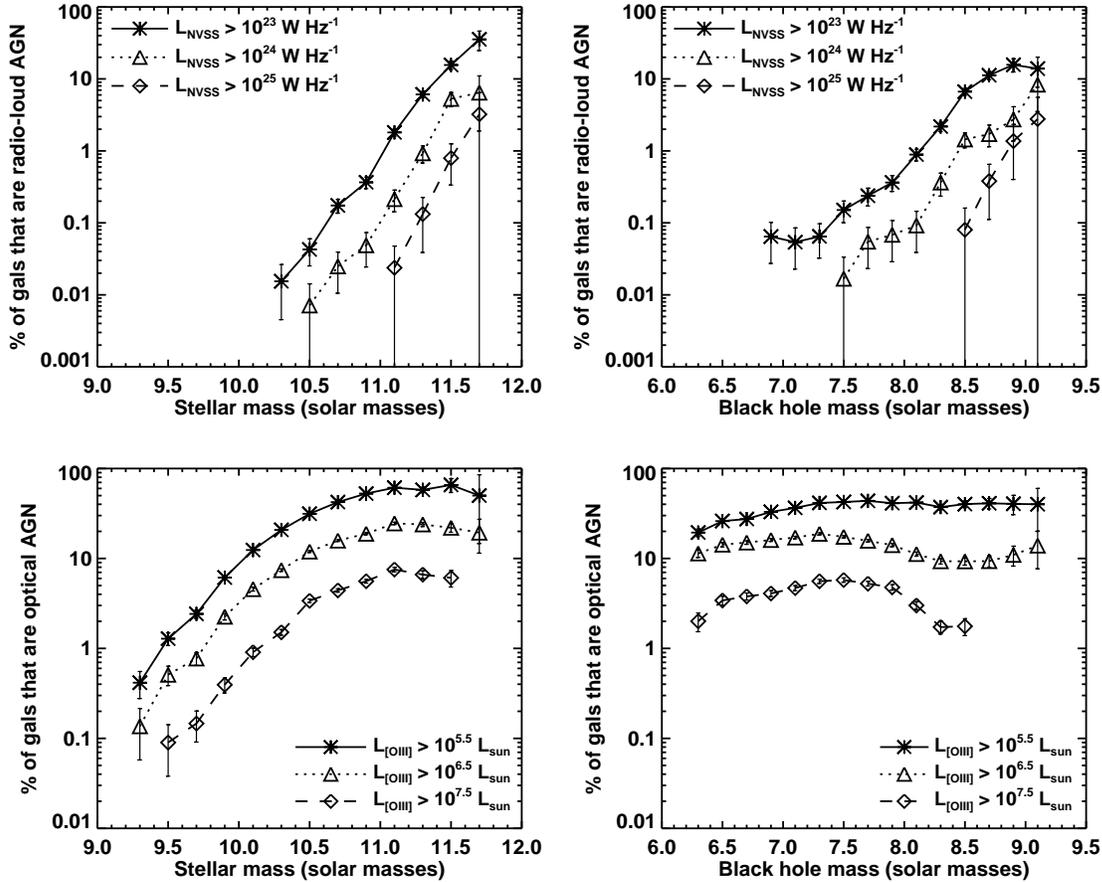,angle=0,width=15cm,clip=}
}
\caption{\label{radloudfrac1} {\it Top:} the fraction of galaxies
which are radio--loud AGN, as a function of stellar mass (left) and black
hole mass (right), for different cuts in radio luminosity. {\it Bottom:}
the equivalent plots for optical emission--line selected AGN. The
radio--loud AGN fraction is a remarkably strong function of stellar or
black hole mass, whilst the optical AGN fraction is largely independent of
black hole mass.}
\end{figure*}

The top--left panel of Figure~\ref{radloudfrac1} shows the fraction of
galaxies with redshifts $0.03 < z < 0.1$ that are classified as
radio--loud AGN (with 1.4\,GHz radio luminosity above
$10^{23}$W\,Hz$^{-1}$), as a function of the stellar mass of the
galaxy. The fraction rises from 0.01\% of galaxies with stellar mass
$3 \times 10^{10} M_{\odot}$ up to over 30\% of galaxies more massive
than $5 \times 10^{11} M_{\odot}$, roughly along a relation $f_{\rm
radio-loud} \propto M_*^{2.5}$.  Also included on this plot
are the radio--loud AGN fractions for radio luminosity cut--offs of
$10^{24}$ and $10^{25}$W\,Hz$^{-1}$. It is notable that all have the
same power--law dependence on stellar mass.

The top right panel of Figure~\ref{radloudfrac1} shows the radio--loud AGN
fraction as a function of black hole mass. This plot is restricted to
bulge--dominated galaxies with surface mass densities above $10^{8.5}
M_{\odot}$kpc$^{-2}$, in order to restrict the sample to galaxies for
which the black hole mass can be reliably determined from the velocity
dispersion (note that the results are not very much affected if this
condition is removed). Once again, a strong trend with mass is seen, but
the slope of the relation is shallower: $f_{\rm radio-loud} \propto M_{\rm
BH}^{1.6}$. Some of this reduction in slope (and in particular the
flattening at the highest black hole masses) may be due to the errors in
the black hole mass estimates, spreading the sources along the x--axis. A
more important effect is that a large fraction of lower mass galaxies
($M_* \lta 10^{11}M_{\odot}$) are disk dominated and so possess only small
black holes. These galaxies host fewer radio--loud AGN and as a result
$f_{\rm radio-loud}$ will have a steeper dependence on stellar mass than
on black hole mass.

\begin{figure}
\centerline{
\psfig{file=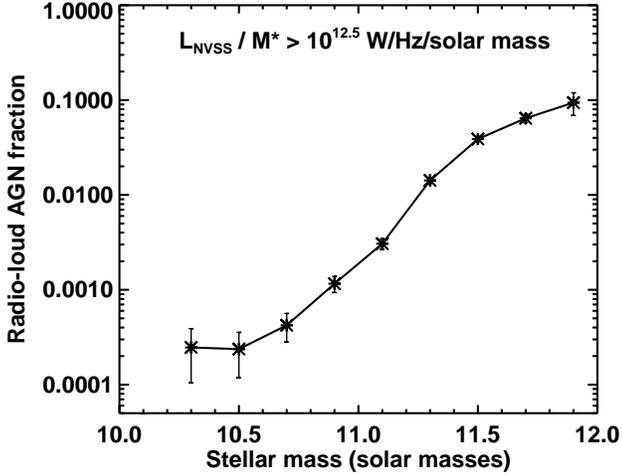,angle=0,width=8.4cm,clip=}
}
\caption{\label{radloudfracedd} The fraction of galaxies, as a function of
stellar mass, that are radio loud AGN, where radio--loud is defined in
terms of being above a given radio luminosity per unit stellar mass (for
bulge--dominated galaxies, this means above a fixed fraction of the
Eddington luminosity). The strong mass dependence remains.}
\end{figure}

It is important to consider whether the strong mass dependence in
Figure~\ref{radloudfrac1} is simply the result of more massive galaxies
having more powerful central engines which may naturally have more
luminous radio emission.  Figure~\ref{radloudfracedd} shows $f_{\rm
radio-loud}$ as a function of stellar mass, where $f_{\rm radio-loud}$ is
now defined to include galaxies brighter than a fixed limit in $L_{\rm
NVSS} / M_*$. For bulge--dominated galaxies, this is equivalent to
defining a galaxy as radio--loud if it is radiating above some fixed
fraction of the Eddington limit. The strong mass dependence remains. The
same result is found if black hole mass rather than stellar mass is used.

In order to ensure that neither aperture nor luminosity selection effects
are influencing the observed results, these studies (and those in later
sections of the paper) have also been repeated using only those galaxies
within fixed narrow ranges in redshift (e.g. $0.06 < z < 0.07$). The same
results are obtained (albeit somewhat noisier), indicating that redshift
effects are not a problem.

\subsection{Optical AGN fractions}

In Figure~\ref{bh_mass}, the emission--line selected AGN have lower black
hole masses on average than radio--loud AGN of the same stellar mass. This
means that these AGN are not preferentially located in bulge--dominated
galaxies. Indeed, their black hole masses even appear to be lower than
those of inactive galaxies, suggesting that optical AGN are found
preferentially in galaxies with substantial disk components.

The dependence of the emission--line selected AGN fraction upon mass is
also strikingly different to that of the radio--loud AGN. The lower panels
of Figure~\ref{radloudfrac1} show this for different cuts in
(extinction--corrected) [OIII]~5007 line luminosity. As for the
radio--loud AGN fractions, the analysis is limited to galaxies with
redshifts $0.03 < z < 0.10$ for the two brighter luminosity cuts; for the
faintest cut the upper redshift limit is reduced to 0.08 in order to
retain completeness in the sample. Optical AGN are preferentially found in
galaxies of high stellar mass, with the proportion of galaxies hosting AGN
with $L_{[OIII]} > 10^{5.5} L_{\odot}$ increasing strongly up to stellar
masses of $10^{10.5} M_{\odot}$ and remaining roughly constant above this
value (cf. K03).\nocite{kau03c} This variation can be largely explained by
the fact that the fraction of disk-dominated galaxies (with only very low
mass black holes) increases at low stellar masses: when plotted against
black hole mass instead of stellar mass, the emission line AGN fraction
shows much less variation.
   
At higher [OIII] luminosities ($L_{[OIII]} > 10^{7.5} L_{\odot}$), the AGN
fraction falls at the highest masses. These results have been discussed in
detail in K03 and Heckman \etal\ \shortcite{hec04} and reflect the fact
that both the presence of a black hole and sufficient gas content are
necessary to achieve high emission line luminosities in
optically--selected AGN. Although massive galaxies host the biggest black
holes, they generally have old stellar populations and very little gas. As
a result they do not usually host powerful emission--line AGN at the
present day.

Radio selection of AGN picks out the most massive galaxies with the
biggest black holes, and it is clear that this is a very different
population of objects to that determined by emission--line selection. The
relation between radio and optical activity is investigated in detail in
Section~\ref{radoptindep}.

\subsection{The bivariate radio luminosity -- mass function}
\label{bivar}

\begin{figure*}
\begin{tabular}{c}
\psfig{file=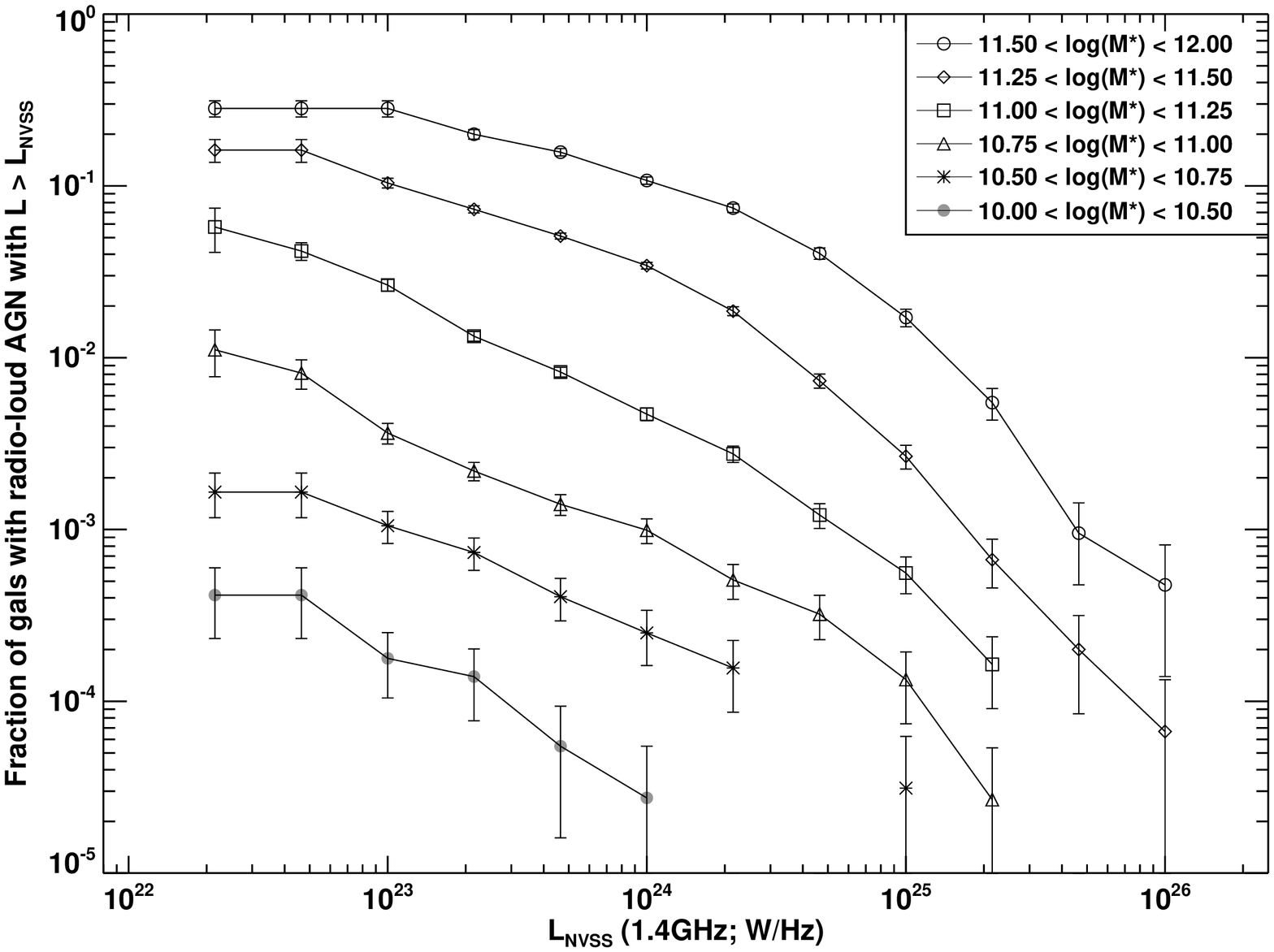,angle=90,width=11cm,height=7.6cm,clip=}
\\
\psfig{file=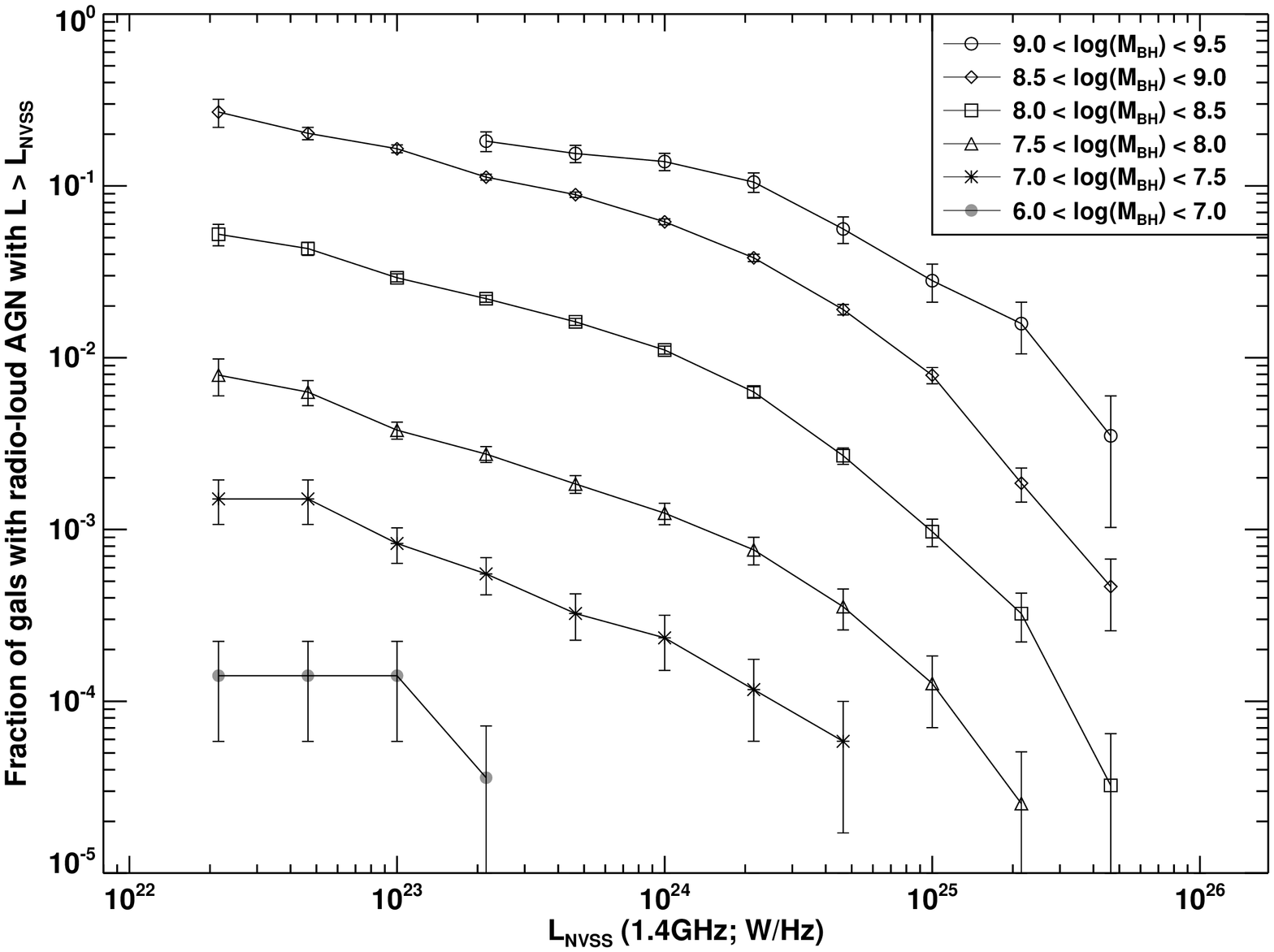,angle=90,width=11cm,height=7.6cm,clip=}
\\
\psfig{file=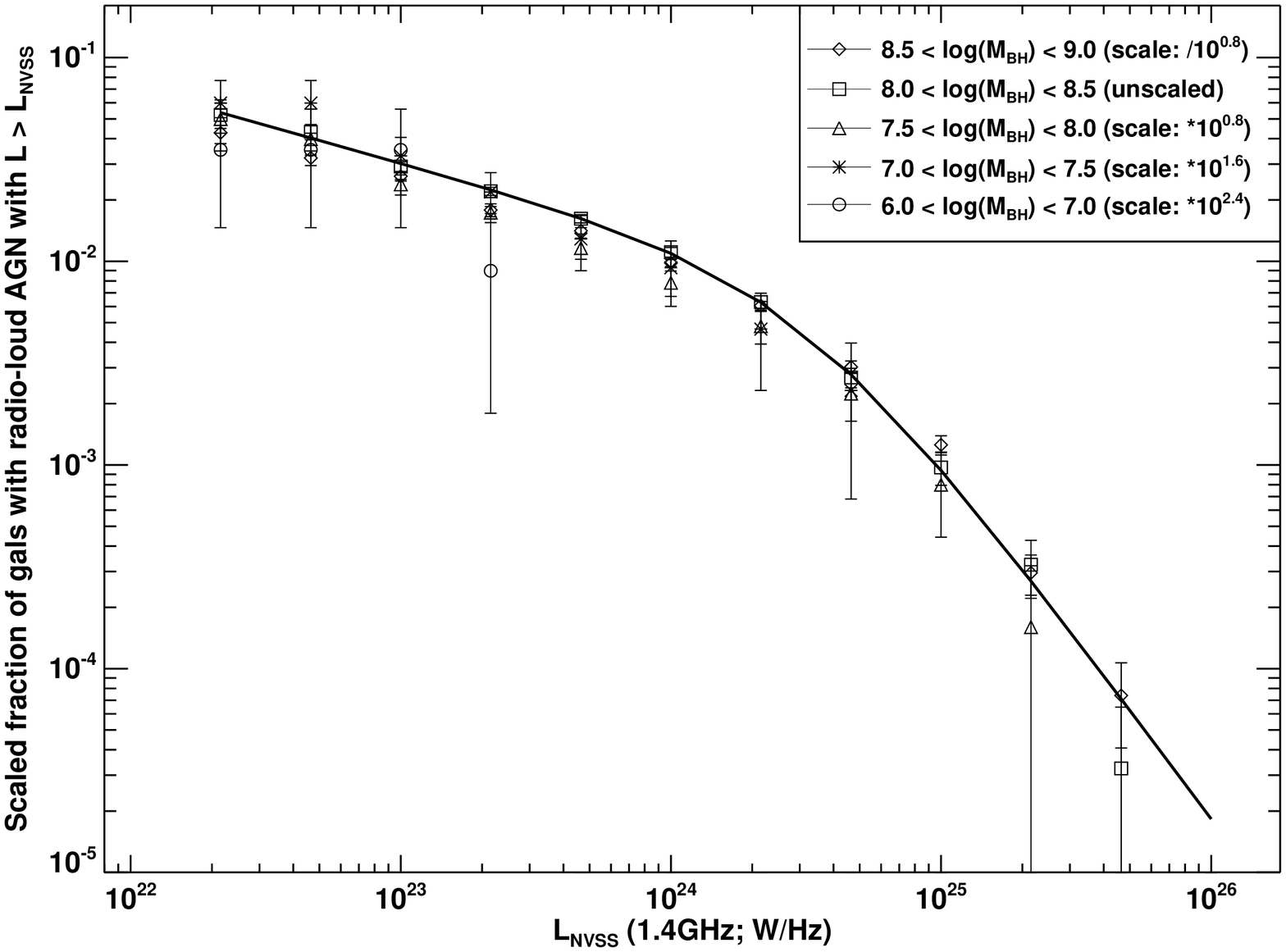,angle=90,width=11cm,height=7.6cm,clip=}
\end{tabular}
\caption{\label{agnfracs} The fraction of radio--loud AGN brighter than a
given radio luminosity, as a function of stellar mass (top panel) and
black hole mass (middle panel). The bottom panel shows the black hole mass
relations scaled by $M_{\rm BH}^{1.6}$ and fitted with a broken power
law. }
\end{figure*}

Determination of the bivariate radio--optical luminosity function was
first carried out by Auriemma \etal\ \shortcite{aur77}, and
investigations using improved samples were carried out by Sadler
\etal\ \shortcite{sad89} and by Ledlow \& Owen \shortcite{led96}.
These authors were able to show that the probability of a galaxy being
radio--loud is a strong function of its optical luminosity. The Sadler
\etal\ and Ledlow \& Owen studies also argued that the location of the
break in the radio luminosity function scaled with optical
luminosity. The uncertainties on these bivariate luminosity functions
were large, owing to the small sizes of the samples available for
study. These studies can now be improved using the much larger SDSS
radio source sample; in addition, the SDSS data allow a comparison
with stellar mass or black hole mass instead of optical luminosity.

Figure~\ref{agnfracs} shows the integral bivariate radio luminosity versus
mass functions derived for the SDSS sample. The top panel shows the
fraction of galaxies which are radio--loud AGN brighter than a given radio
luminosity, as a function of radio luminosity. Results have been plotted
for six bins in stellar masses. The middle panel shows the equivalent
relations for six bins in black hole mass. The most striking feature of
these figures is the dramatic decrease in the fraction of radio--loud AGN
with decreasing stellar or black hole mass, as discussed above. Another
remarkable feature is that for both stellar and black hole mass the shape
of the functions are very similar for all mass ranges (except for the
highest mass bins, discussed below). This is illustrated in the bottom
panel of Figure~\ref{agnfracs}, where the integral bivariate radio
luminosity functions for black hole mass have been scaled by $M_{\rm
BH}^{1.6}$ to remove the mass dependence. The scaled data agree very well,
and there is no evidence for any dependence of the break luminosity on
black hole mass. These results indicate that the earlier suggestions of
Sadler \etal\ \shortcite{sad89} and Ledlow \& Owen \shortcite{led96} of a
dependence of the break radio luminosity on the optical luminosity of the
host galaxy were either artefacts of the small sample sizes or differences
resulting from the use of optical luminosity instead of black hole mass.

Ledlow \& Owen found evidence for a flattening of the faint end slope of
the radio luminosity function with increasing optical luminosity of the
host galaxy. These results are not supported by the current data except at
the highest stellar and black hole masses ($M_* \gta 10^{11.5} M_{\odot}$,
$M_{\rm BH} \gta 10^9 M_{\odot}$) where there is evidence for shallower
slopes. It is likely that the Ledlow \& Owen result was entirely driven by
these highest mass sources. The precise physical explanation of the
flattening of the faint--end slope at the highest masses is not yet known,
but what is clear is that a turn--over would have to occur at some mass,
otherwise the radio--loud fraction would exceed 100\%.

In summary, Figure~\ref{agnfracs} demonstrates that the {\it probability}
of a galaxy becoming a radio source is a very strong function of its
mass. Except at the very highest masses, however, the {\it luminosity} of
the radio source that results is independent of mass. The bottom panel of
Figure~\ref{agnfracs} shows that this luminosity dependence can be well
fitted by a broken power law model such that overall the fraction of
sources that are radio loud AGN brighter than some luminosity $L$ is given
by:

\begin{displaymath}
f_{\rm radio-loud} = f_0 \left(\frac{M_{\rm BH}}{10^8 M_{\odot}}\right)^{\alpha}
\left[\left(\frac{L}{L_*}\right)^{\beta} +
\left(\frac{L}{L_*}\right)^{\gamma}\right]^{-1}
\end{displaymath}

\noindent for which the best--fit parameter values are as follows:

\begin{displaymath}
\begin{array}{rcl}
f_0 & = & (3.5 \pm 0.4) \times 10^{-3}\\
\alpha & = & 1.6 \pm 0.1\\
\beta & = & 0.37 \pm 0.03\\
\gamma & = & 1.79 \pm 0.14\\
L_* & = & (3.2 \pm 0.5) \times 10^{24} {\rm W Hz}^{-1}
\end{array}
\end{displaymath}

This equation holds well for all probabilities $f_{\rm radio-loud} \lta
0.25$, but breaks down at higher values of $f_{\rm radio-loud}$, possibly
reflecting a ceiling at the $\sim 30$\% level.

A similar relation can be derived as a function of stellar mass (in units
of $10^{11} M_{\odot}$), and provides best--fit values of $f_0 = 0.0055
\pm 0.0004$, $\alpha = 2.5 \pm 0.2$, $\beta=0.35\pm0.03$, $\gamma=1.54\pm
0.11$ and $L_* = (2.5 \pm 0.4) \times 10^{24}$W\,Hz$^{-1}$.

\begin{figure}
\centerline{
\psfig{file=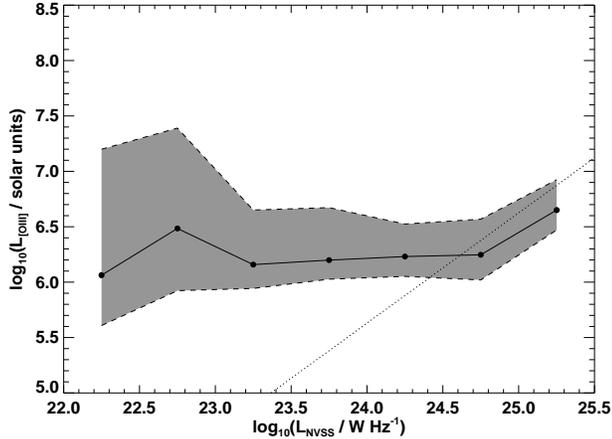,angle=90,width=8.6cm,clip=}
}
\caption{\label{o3lumrad} The median [OIII]~5007 emission line luminosity
of the radio--loud AGN (together with the 25\% and 75\% quartile ranges,
indicated by the shaded regions), as a function of radio luminosity, for
the SDSS radio--loud AGN in the redshift range $0.03 \le z \le
0.10$. There is no evidence for any correlation between the two. The
dotted line shows an extrapolation of the correlation found between these
two parameters for high power (FR\,II) radio galaxies (e.g. Rawlings \&
Saunders 1991, McCarthy 1993, Zirbel \& Baum 1995).}
\end{figure}
\nocite{raw91b,mcc93,zir95}

\section{Independence of radio and optical AGN activity}
\label{radoptindep}

In Paper I it was shown that some radio--loud AGN are classified as
optical AGN based upon their emission line properties, whilst others are
optically inactive. It was also shown that out to redshifts $z \sim 0.1$
the relative numbers of radio--loud AGN which are or are not classified as
emission line AGN are roughly similar: of the 420 radio--loud AGN in the
redshift range $0.03 \le z \le 0.10$, 227 are classified as optical AGN
based upon their emission lines and 193 classified as optically
inactive\footnote{In fact, just over half of the `inactive' galaxies do
have marginal [OIII]~5007 detections, but are not classified as
emission--line AGN because that classification requires all four emission
lines in the emission line diagnostic diagram to be detected with
S/N $>$ 3 (cf. K03; Paper I).}. At higher redshifts
the proportion of emission line AGN decreases rapidly, because weak
nuclear emission lines become increasingly more difficult to detect. In
this section, the relationship between the radio and emission line
activity is investigated.

It is well-known that for very powerful radio sources, emission line and
radio luminosity are well--correlated (e.g.  Rawlings \& Saunders 1991,
McCarthy 1993). However, the SDSS radio sources are less powerful than
these, and Zirbel \& Baum \shortcite{zir95} have shown that emission line
luminosity depends much less strongly upon radio luminosity at these lower
radio luminosities. Figure~\ref{o3lumrad} shows the median
extinction--corrected [OIII]~5007 emission line luminosity as a function
of radio luminosity (corrected for the contribution due to star formation,
as described in Paper I) for the SDSS radio--loud AGN in the redshift
range $0.03 \le z \le 0.10$: no correlation at all is seen between the two
quantities out to a 1.4\,GHz radio luminosity of $10^{25}$W\,Hz$^{-1}$
(the highest luminosity available in this redshift range). This is
consistent with the previous results, and indeed the turn--up in the last
bin may be related to the onset of the correlation found at higher radio
luminosities. It is interesting that the radio luminosity at which the
transition in emission--line properties properties occurs is similar to
that where the transition from low luminosity FR\,Is to the more powerful
FR\,IIs takes place (cf. Baum \etal\ 1995, Zirbel \& Baum 1995, Wills
\etal\ 2004),\nocite{bau95,zir95,wil04} and is also close to the break in
the radio luminosity function.

\begin{figure}
\centerline{
\psfig{file=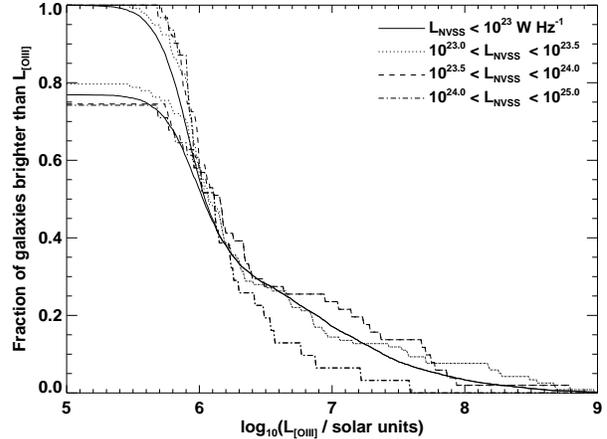,angle=90,width=8.6cm,clip=}
}
\caption{\label{cumo3lum} The fraction of galaxies, with masses in the
narrow range $10^{11} < M/M_{\odot} < 10^{11.5}$, whose [OIII]~5007
emission line luminosity is brighter than a given luminosity, as a
function of that luminosity. This is shown for four different samples of
galaxies: the solid lines show the result for all of the galaxies not
classified as radio--loud AGN, and the dotted, dashed, and dot--dashed
lines show radio--loud AGN for three different ranges in radio
luminosity. At low [OIII]~5007 luminosities each line splits into two:
these two lines represent the uncertainty on the distribution caused by
the presence of galaxies with upper-limits on their [OIII]~5007
luminosity: the lower line of each pair indicates the distribution if all
of the undetected galaxies have zero emission line luminosity, and the
upper line represents the case where each undetected galaxy has an
emission line luminosity equal to the upper limit. There is no
statistically significant difference between any of the four curves.}
\end{figure}

\begin{figure*}
\centerline{
\psfig{file=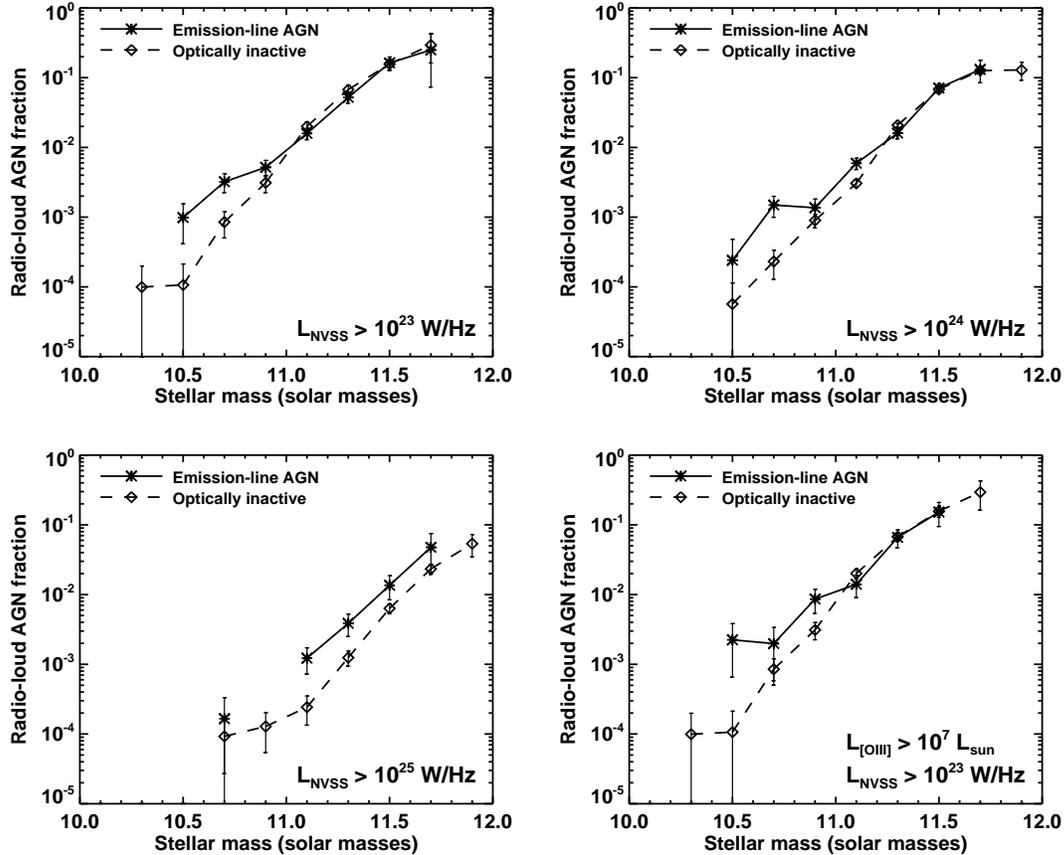,angle=0,width=14.6cm,clip=}
}
\caption{\label{radloudfrac2} The fraction of emission--line selected AGN
(solid line) and optically--inactive galaxies (dashed line) which are
classified as radio--loud AGN, as a function of stellar mass. The
upper--left panel shows the relations for those radio--loud AGN with NVSS
radio luminosities brighter than $10^{23}$W\,Hz$^{-1}$. The near-identical
nature of these relations indicates that radio--loud AGN activity is
essentially independent of whether or not a source is optically classified
as an AGN.  The upper--right panel shows that the general form of the
result is insensitive to increasing the radio luminosity limit to
$10^{24}$W\,Hz$^{-1}$ (although the specific fractions change), but the
lower--left panel demonstrates that differences begin to be seen above
$10^{25}$W\,Hz$^{-1}$. The lower--right panel demonstrates that the result
is unchanged if the classification of optical AGN is restricted only to
the most emission--line luminous. }
\end{figure*}

Figure~\ref{cumo3lum} investigates this result in more detail, comparing
the distribution of [OIII]~5007 emission line luminosities for samples of
radio--loud AGN in three different ranges of radio luminosity with a
`control' sample of galaxies that are not detected in the radio. The
analysis is restricted to galaxies with stellar masses in the range
$10^{11} < M/M_{\odot} < 10^{11.5}$; a narrow mass range is adopted to
avoid biases associated with the strong increase in the fraction of radio
galaxies as a function of stellar mass.  The samples are also restricted
to early-type galaxies with surface mass densities above $10^{8.5}
M_{\odot}$kpc$^{-2}$ (this additional criterion does not remove any AGN
and only excludes 1.5\% of the galaxies in the control sample, so has
negligible effect on the final result). A number of the galaxies only have
upper limits to their [OIII]~5007 luminosities. These are accounted for by
plotting two lines for each sample, representing the upper and lower
bounds of the cumulative distribution (see figure caption for more
details). It is apparent that there are no major differences between the
samples of radio--loud AGN and the control sample. The {\sc ASURV}
survival analysis package \cite{lav92} was used to confirm this
result. This calculates the probability that two samples are drawn from
the same parent distribution according to 5 different survival analysis
methods (which take proper account of the upper limits), each based around
a generalised Wilcoxon test. For each of the three radio luminosity bins,
the mean probability of drawing the radio--loud AGN sample out of the same
parent population as the control sample of radio--quiet galaxies was
greater than 25\%. This proves that, at least for galaxies with these
stellar masses, the emission line luminosity is independent of radio
luminosity.

It is important to ascertain whether this result holds for all stellar
masses and radio luminosities. The upper--left panel of
Figure~\ref{radloudfrac2} shows the fraction of galaxies that are
radio--loud AGN, as a function of stellar mass, for two different galaxy
subsamples. The solid line shows $f_{\rm radio-loud}$ for those galaxies
classified as AGN by their emission--line properties. The dashed line
shows the same thing for optically inactive galaxies.  It is remarkable
that not only do these relations show broadly the same mass dependence,
but also the values of $f_{\rm radio-loud}$ for the two subsamples agree
almost perfectly at the highest masses. At stellar masses greater than
$10^{11} M_{\odot}$, the probability that a galaxy is radio--loud is
independent of whether it is classified as an emission--line AGN. This
confirms that (low radio luminosity) radio and emission line AGN activity
are two physically distinct and independent phenomena. At lower stellar
masses the two lines appear to diverge, although a larger sample would be
preferable to confirm that this result is statistically significant. If it
is, it might indicate that at lower masses there is more overlap between
the radio--loud and emission line AGN samples, possibly due to a
population of radio--loud Seyferts galaxies where the production of jets
and ionising radiation are linked (e.g. De Bruyn \& Wilson 1978; Xu \etal\
1999).\nocite{deb78,xu99}

The independence of radio and optical activity for high mass galaxies
still holds if the radio luminosity limit is raised to
$10^{24}$W\,Hz$^{-1}$ (upper--right panel of
Figure~\ref{radloudfrac2}). Likewise the result remains unchanged if the
comparison is restricted to optically powerful AGN, with emission line
strengths above $10^7 L_{\odot}$ (lower--right panel of
Figure~\ref{radloudfrac2}). A small difference is seen if the radio
luminosity limit is set as high as $10^{25}$W\,Hz$^{-1}$ (lower--left
panel of Figure~\ref{radloudfrac2}): the fraction of radio--loud AGN is
then higher amongst optical AGN than amongst optically inactive galaxies.
This is probably a consequence of the correlation between radio and
emission line luminosities that sets in at high radio luminosities.

\begin{figure*}
\centerline{
\psfig{file=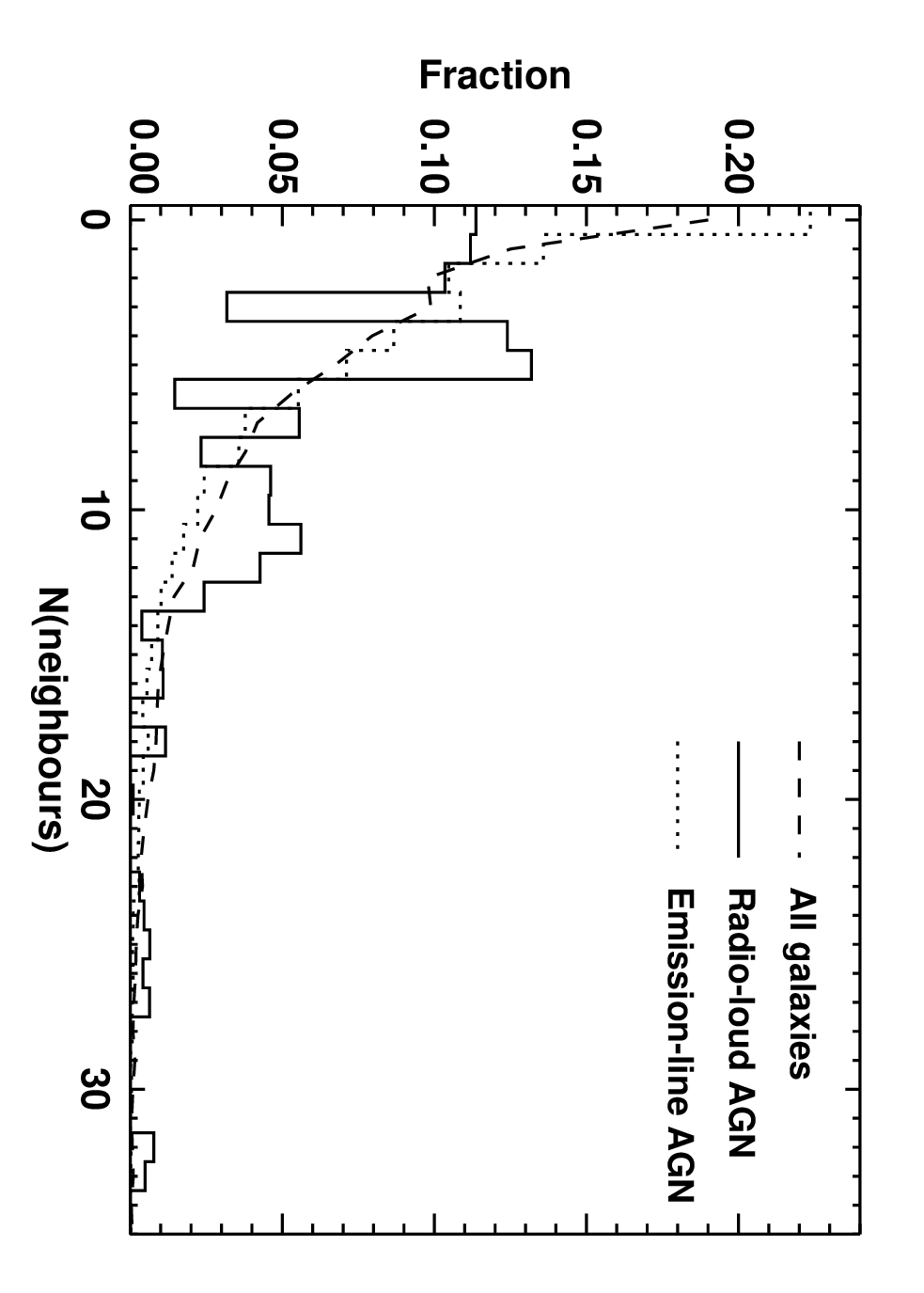,angle=90,width=9cm,clip=}
}
\centerline{
\psfig{file=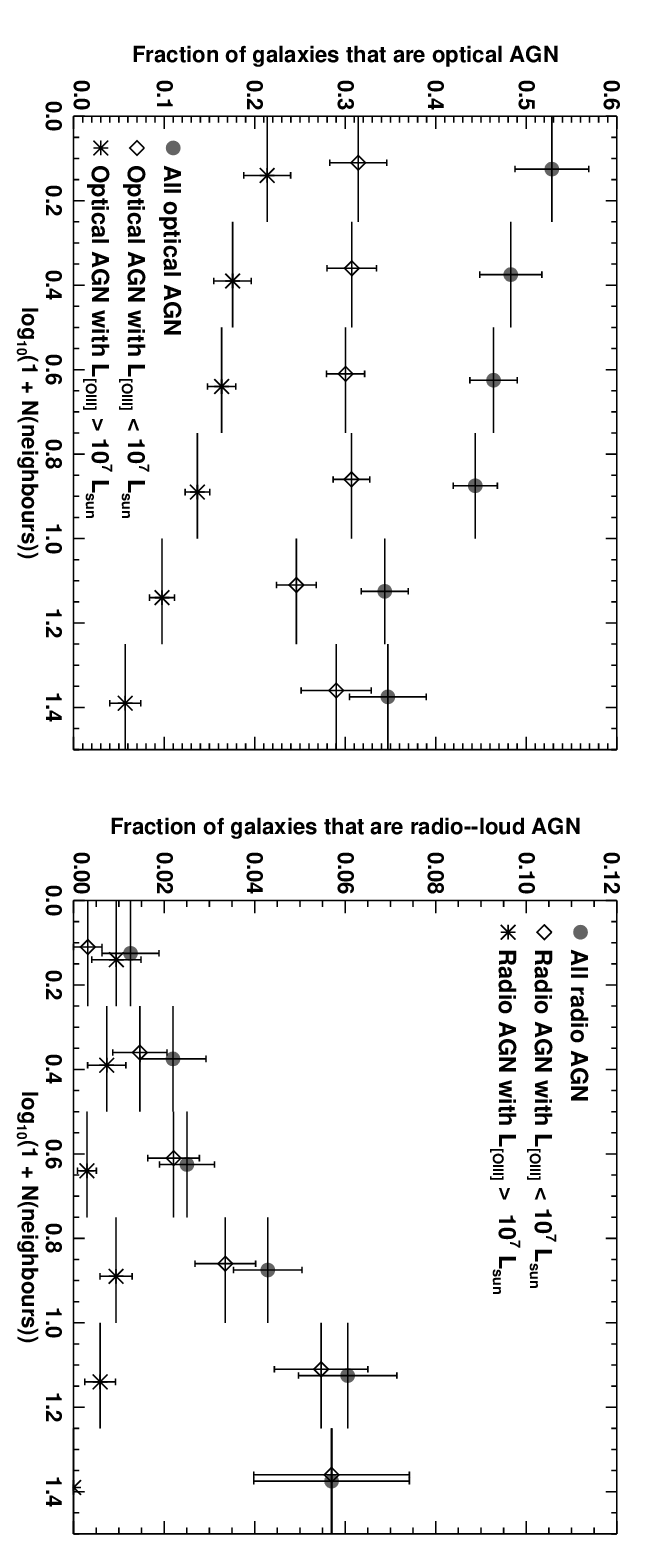,angle=90,width=16cm,clip=}
}
\caption{\label{o3env} {\it Top:} The distribution of local environments
of radio--loud AGN, emission--line selected AGN, and all galaxies, for
galaxies with stellar masses $M_* > 10^{10} M_{\odot}$. The $N$ parameter
represents the number of galaxies above a fixed absolute magnitude limit
that lie within a 2\,Mpc projected radius and within a velocity of $\pm
500$\,km\,s$^{-1}$ of the galaxy under study (cf. Kauffmann \etal\ 2004).
{\it Bottom-left:} The fraction of galaxies with masses between $10^{11}$
and $5 \times 10^{11} M_{\odot}$ which host emission--line selected AGN, as a
function of local environmental density. Also shown are the distributions
of the two sub-samples of this with extinction corrected [OIII]
emission--line luminosities $L < 10^7 L_{\odot}$ and $L > 10^7 L_{\odot}$
(these points have each been offset slightly along the x--axis for
clarity). The fraction of luminous emission line AGN falls dramatically
with increasing environmental density, but that of low luminosity AGN show
no environmental dependence (cf. Kauffmann \etal\ 2004). {\it
Bottom-right:} An equivalent plot of the environmental dependence of
radio--loud AGN activity, as a function of emission--line
luminosity. Except at the highest emission--line luminosities, the
fraction of galaxies hosting radio--loud AGN increases towards richer
environments.}
\end{figure*}

\section{Radio and optical AGN activity as a function of environment}
\label{envdep}

Best \shortcite{bes04a} studied the environmental dependence of
radio--loud AGN activity in the 2dFGRS. The fraction of radio--loud AGN
was found to be largely independent of local galaxy density (as evaluated
by the distance to the tenth nearest neighbour), although a dependence was
found on the larger scale cluster\,/\,group\,/\,field environment.
Further, when the 2dFGRS radio--loud AGN population was separated into
subsets of emission--line AGN and optically inactive AGN, it was apparent
that the two classes had different environmental dependences.  In
particular, the fraction of galaxies hosting radio--loud AGN with no
emission lines {\it increased} towards richer environments.

The SDSS sample allows a detailed investigation of these results.
Kauffmann \etal\ \shortcite{kau04} estimated the local environmental
density of SDSS galaxies by deriving the number of galaxies ($N$) above a
fixed absolute magnitude limit, within a 2\,Mpc projected radius and a
velocity of $\pm 500$\,km\,s$^{-1}$ of the galaxy under study. The top
panel of Figure~\ref{o3env} shows the distribution of all galaxies,
emission--line selected AGN, and radio--loud AGN brighter than
$10^{23}$W\,Hz$^{-1}$, as a function of environmental parameter $N$. To
devire these distributions each galaxy has been weighted by the inverse of
the volume in which it could have been detected (see K03 for a full
discussion of how these were derived), and analysis has been restricted to
only galaxies with stellar mass above $10^{10} M_{\odot}$, and redshift
$0.03 < z < 0.1$.

\begin{figure*}
\centerline{
\psfig{file=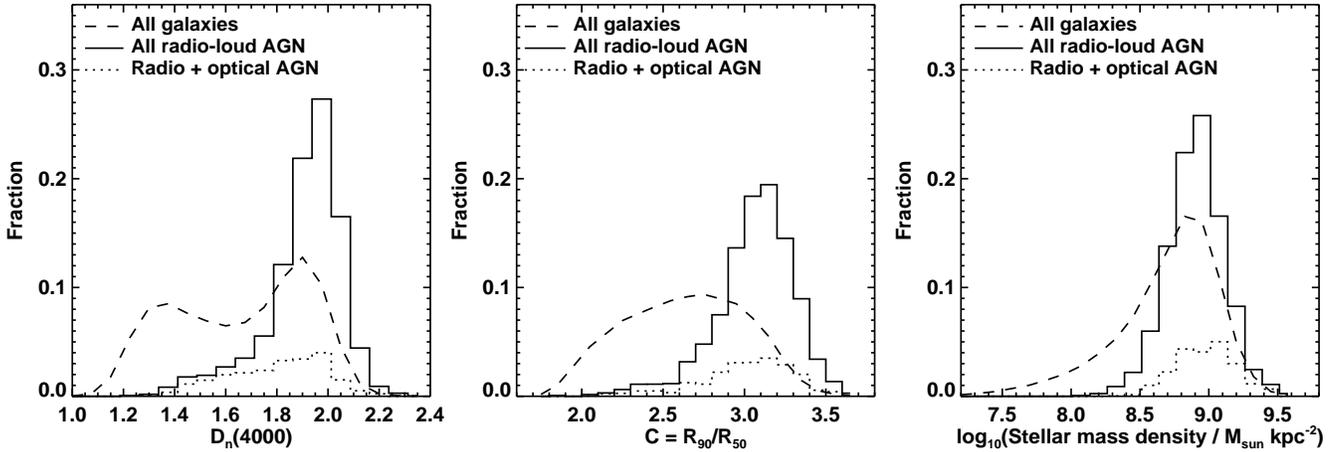,angle=90,width=\textwidth,clip=}
}
\caption{\label{prophists} The dashed lines show the fraction of total
stellar mass in the nearby Universe ($z < 0.1$) contained in galaxies as a
function of 4000\AA\ break strength (left), concentration index
($C=R_{90}/R_{50}$; middle) and stellar surface mass density $\mu_* = M_*
/ 2\pi R_{50}^2$; right). The solid histograms show the equivalent
relations for radio--loud AGN, and the dotted histograms show the subset
of these that also display emission--line AGN activity.}
\end{figure*}

As found by Kauffmann \etal\ \shortcite{kau04} and also by Miller \etal\
\shortcite{mil03a}, the distribution of environments of the emission--line
selected AGN is similar to that of normal galaxies.  Kauffmann \etal\
showed that the fraction of low emission--line luminosity AGN (LINERS) is
essentially independent of local galaxy density, but that for high
emission--line luminosity AGN (mostly Seyferts), the AGN fraction is found
to fall in regions of high galaxy density. These results are clearly
demonstrated in the lower--left panel of Figure~\ref{o3env}; this shows,
for galaxies in the narrow range of stellar masses $10^{11} < M_* /
M_{\odot} < 5 \times 10^{11}$, the overall fraction of emission--line AGN
as a function of environmental density, together with the split of this
into high and low [OIII]~5007 emission line luminosity subsamples. There
is only an environmental dependence for AGN with the highest emission line
luminosities ($L_{\rm [OIII]} > 10^7 L_{\odot}$).

The radio--loud AGN in the top panel of Figure~\ref{o3env} clearly favour
denser environments than normal galaxies, with a much longer tail towards
the richest environments. The lower--right panel of Figure~\ref{o3env}
shows that the radio--loud AGN fraction increases significantly as a function
of environment. Once again, large differences are seen if the sample is
split by [OIII] emission line luminosity: those radio--loud AGN with the
strongest emission lines (L[OIII]$>10^7 L_{\odot}$) preferentially avoid
the densest regions.

This result indicates that environment, as well as black hole mass, may be
an additional factor in determining whether a galaxy becomes a radio--loud
AGN. A full investigation of environmental effects is beyond the scope of
the current paper; a detailed study of these will be carried out in the
third paper of this series (Kauffmann \etal\ 2005, in prep.)

\section{Host galaxy properties of radio--loud AGN}
\label{hostprops}

In earlier sections of this paper it was shown that there is a strong
dependence of the radio--loud AGN fraction on the stellar or black hole
mass of the galaxy. In addition, at a given stellar mass, radio--loud AGN
are biased towards galaxies with more massive bulges. In this Section, the
host galaxies of radio--loud AGN are compared with those of emission--line
selected AGN and inactive galaxies. The host galaxy properties that are
considered are the 4000\AA\ break strength $D_n(4000)$, the concentration
index $C$, defined as $C=R_{90}/R_{50}$ where $R_{90}$ and $R_{50}$ are
the radii containing 90\% and 50\% of the $r$-band light of the galaxy,
and the surface mass density $\mu_* = M_* / 2\pi R_{50}^2$. The 4000\AA\
break strength provides a measure of the mean stellar age of the galaxy
(cf. K03) and the $C$ and $\mu_*$ parameters allow investigation of its
structure. Analysis is restricted to galaxies with $z < 0.1$, in order
that the host galaxy parameters can be reliably determined.

\begin{figure*}
\centerline{
\psfig{file=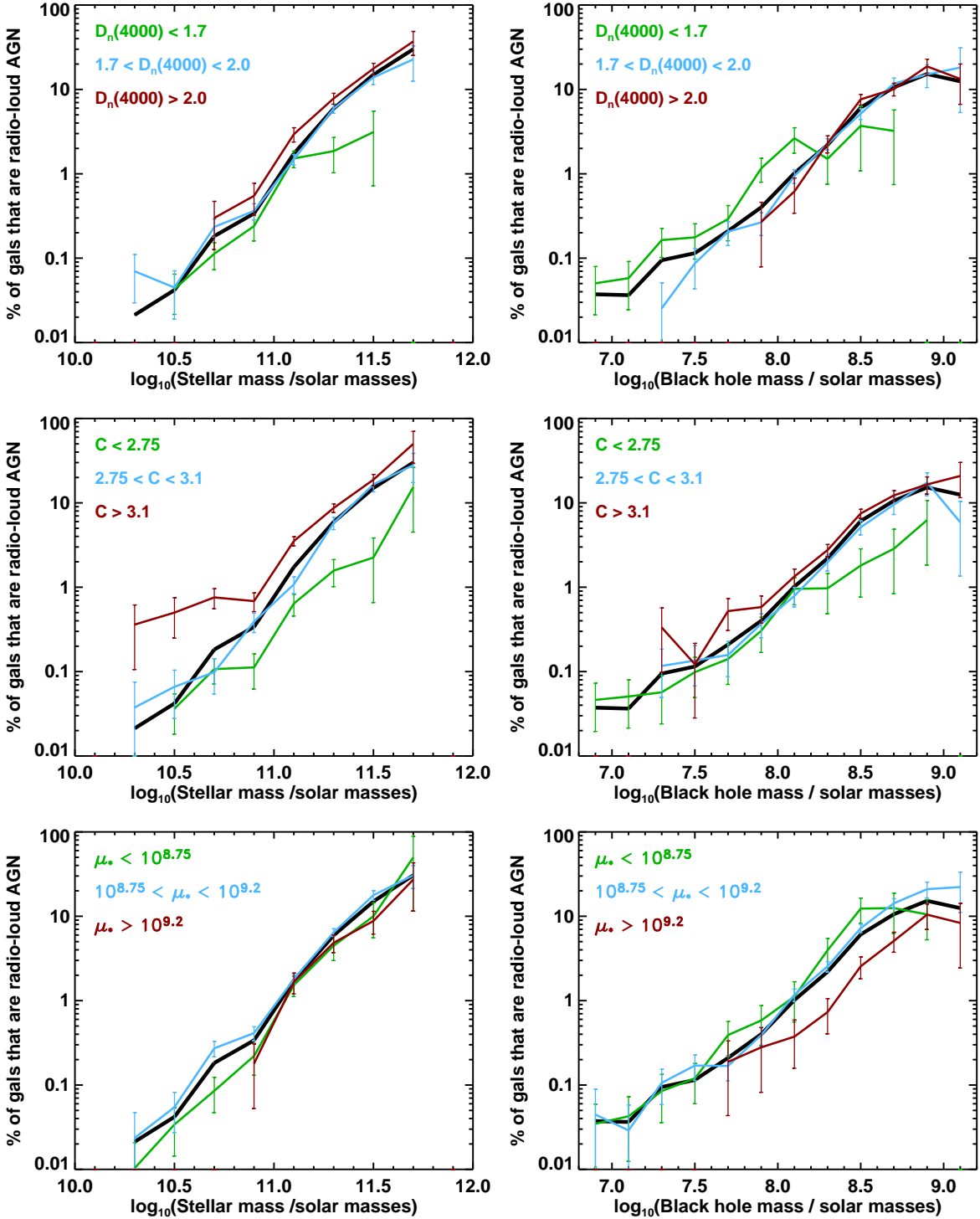,angle=0,width=16cm,clip=} 
}
\caption{\label{galpropsfig} As a function of stellar mass (left) or black
hole mass (right), the thick solid line shows the proportion of galaxies
which are radio--loud AGN (cf. Figure~\ref{radloudfrac1}), and the other
three lines show the equivalent fractions for galaxies with a specific
range of host galaxy properties {\it Top:} 4000\AA\ break strength; {\it
Middle:} concentration index; {\it Bottom:} stellar surface mass
density. }
\end{figure*}

Figure~\ref{prophists} compares the distribution of these parameters for
all galaxies and for radio--loud AGN brighter than $10^{23}$W\,Hz$^{-1}$
at 1.4\,GHz. Radio--loud AGN typically have large 4000\AA\ break
strengths, high concentration indices, and moderate--to--high stellar
surface mass densities. None of these is particularly surprising since all
of these parameters are tightly correlated with stellar mass
(e.g. Kauffmann \etal\ 2003b)\nocite{kau03a}. In order to remove this
stellar mass dependence, Figure~\ref{galpropsfig} shows the proportion of
galaxies that are radio loud AGN as a function of both stellar and black
hole mass, for three different bins of 4000\AA\ break strength (top
panels), concentration index (middle panels) and surface mass density
(bottom panels).

At fixed stellar mass, $f_{\rm radio-loud}$ is higher, at all stellar
masses, for larger values of D$_n$(4000) and $C$. This is another
manifestation of the effect shown in Figure~\ref{bh_mass}: at a given
stellar mass, radio galaxies are preferentially located in galaxies with
larger bulge-to-disk ratios, and these typically have older stellar
populations and higher concentration indices.  The dependence of $f_{\rm
radio-loud}$ on D$_n$(4000) and $C$ is much weaker at fixed black hole
mass (i.e.  fixed velocity dispersion $\sigma$), which also supports this
picture. Indeed, $f_{\rm radio-loud}$ shows little dependence on either
$D_n(4000)$ or $C$ in the upper two bins, and only differs for the
(typically more disky) galaxies in the bins with $D_n(4000) < 1.7$ and $C
< 2.75$. This may be related to the secondary dependence of $f_{\rm
radio-loud}$ on environment discussed in the previous section: radio--loud
AGN are preferentially located in rich environments and galaxies in
high-density regions of the Universe tend to have older stellar
populations and larger bulge-to-disk ratios.

The lower--right panel of Figure~\ref{galpropsfig} indicates that at fixed
black hole mass, $f_{\rm radio-loud}$ is higher for galaxies with lower
stellar surface mass densities. Large sizes (low $\mu_*$), as well as
extended envelopes (high $C$), appear to increase the probability of
galaxies being radio--loud. Interestingly, these properties are typical of
the central galaxies of groups and clusters, and it is well-known that the
central galaxies of groups and clusters are frequently radio--loud AGN
(e.g. Burns 1990)\nocite{bur90}. A higher radio--loud fraction for
galaxies with this special location may therefore be an important factor
in this result. However, given the relatively small number of such
galaxies, it is not clear that this effect alone would be sufficient. An
alternative possibility is that this is because if the radio lobes are
confined by a dense medium then adiabatic expansion losses are reduced and
consequently a radio source of given jet power is more radio luminous
(cf. Barthel \& Arnaud 1996, and references therein)\nocite{bar96a}. For
galaxies with lower $\mu_*$ and higher $C$, the interstellar medium is
spread over a larger physical radius (and may also be denser since, at
least for spirals, galaxies with lower stellar surface mass densities are
known to have have higher gas fractions; McGaugh \& de Blok
1997)\nocite{mcg97}.  This means that an expanding radio source will be
located in a denser environment for a longer period of time, which may
boost the average radio luminosities (and possibly the lifetimes) of the
sources located in these systems. In order to account for a factor 2--3
difference in radio--loud AGN fraction between galaxies with $\mu_* \sim
10^{8.75}$ and $\mu_* \sim 10^{9.2}$, radio luminosities would need to be
higher in the lower surface density galaxies by a factor of 3--4, or
lifetimes longer by a factor 2--3. Detailed modelling is required to
confirm whether such factors could be achieved.

Finally, it is interesting to note that in Section~\ref{radoptindep} it
was demonstrated that radio and emission--line AGN activity are
independent. It thus follows that radio--loud AGN with emission lines
should be located in the same host galaxies as their radio--quiet
counterparts. K03 found that emission--line AGN had similar structural
parameters to inactive galaxies of the same stellar mass, but that the
most optically luminous AGN had significantly lower 4000\AA\ breaks.  The
dotted lines on Figure~\ref{prophists} show the contribution to the
distribution of radio--loud AGN made by those that are also emission--line
selected AGN. In both $C$ and $\mu_*$ these have very similar
distributions to those of radio--loud AGN as a whole, but it is
interesting to note that the long tail towards lower values of the
$D_n(4000)$ distribution is almost entirely composed of these
emission--line AGN. The fraction of galaxies that are emission--line
radio--loud AGN thus exhibits little dependence on $\mu_*$ or $C$, but an
increase towards lower $D_n(4000)$, exactly as found by K03 for
radio--quiet objects.

\section{Discussion and Conclusions} 
\label{discuss}

The main results of this paper are as follows:

\begin{itemize}
\item The {\em fraction} of galaxies which host radio--loud AGN is a
remarkably strong function of both stellar mass and black hole mass (as
measured by the central stellar velocity dispersion).  $f_{\rm
radio-loud}$ scales as $M_*^{2.5}$ and $M_{\rm BH}^{1.6}$.  This is in
sharp contrast to the behaviour of the emission-line AGN fraction, which
is largely independent of black hole mass.

\item The {\em distribution} of radio luminosities does not generally
depend on black hole mass. The integral bivariate radio luminosity versus
black hole mass function can be well approximated by a single function,
with the fraction of galaxies which host radio--loud AGN brighter than
luminosity $L$ scaling as $f_{\rm radio-loud} = f_0 (M_{\rm BH}/10^8
M_{\odot})^{\alpha} [(L/L_*)^{\beta} + (L/L_*)^{\gamma}]^{-1}$ with
parameter values of $f_0 = 0.0035 \pm 0.0004$, $\alpha = 1.6 \pm 0.1$,
$\beta = 0.37 \pm 0.03$, $\gamma = 1.79 \pm 0.14$ and $L_* = (3.2 \pm 0.5)
\times 10^{24}$W\,Hz$^{-1}$. This function only breaks down when $f_{\rm
radio-loud}$ approaches 20--30\%.

\item Within the range of radio luminosities studied in this paper, radio
and emission-line AGN activity are {\em independent} of each other.  At
fixed high stellar masses, the probability that galaxy is a radio--loud
AGN is not sensitive to whether it classified as an emission-line AGN.

\item At a fixed black hole mass, the host galaxies of radio--loud AGN are
preferentially found in galaxies with higher concentration indices and
lower surface mass densities, together with a possible weak trend to avoid
the lowest 4000\AA\ break strengths. Radio--loud AGN are also
preferentially found within richer environments. It is possible that all
of these results could be related to higher radio luminosities that result
when the radio lobes are confined by denser gas.
\end{itemize}

A number of these results are broadly in line with what might have been
expected, but others deserve some consideration. 

\subsection{Independence of radio and emission line AGN activity} 

The current study indicates that radio and emission line AGN activity are
entirely independent at the radio luminosities studied; at higher radio
luminosities it is well--known that radio and emission line luminosity are
strongly correlated. It is probably a coincidence that the correlation
begins exactly where the current data runs out of sources.  However, it is
interesting to note that at about this radio luminosity, $L_{\rm 1.4GHz}
\approx 10^{25}$W\,Hz$^{-1}$, a fundamental transition occurs in the
properties of radio sources. Above this radio luminosity most sources are
FR\,II radio sources, whilst below that luminosity most are FR\,Is. This
luminosity also corresponds roughly to the break ($L_*$) in the radio
luminosity function. 

The reason for the difference in radio properties between FR\,I and FR\,II
sources is still a matter of much debate. One popular interpretation is
that it may be due to qualitative differences in the properties of the
central black hole, such as the accretion rate, accretion mode, or
black--hole spin (e.g. Baum, Zirbel \& O'Dea 1995)\nocite{bau95}.
Alternatively, interactions with the host galaxy interstellar medium may
be the dominant effect, such that for any given host galaxy only radio
jets above a certain luminosity would be able to bore their way out
through the interstellar medium without disruption (see Snellen \& Best
2001 for a full discussion of the FR\,I--FR\,II dichotomy). Whatever the
origin of this difference, the results of this paper suggest that for
FR\,I type radio sources which dominate the SDSS radio source sample there
is almost no correlation between radio and emission line luminosity. This
indicates that there is a major difference in the way that these radio
sources give rise to their line emission as compared to the more powerful
FR\,IIs (see also Baum \etal\ 1995, Zirbel \& Baum
1995)\nocite{bau95,zir95}.

It is unfortunate that the SDSS radio source sample does not contain many
powerful FR\,II sources. FR\,II radio sources typically have luminous high
ionisation emission line regions, and it would be very interesting to see
whether their host galaxy properties are similar to those of other
emission--line selected AGN. Indications from earlier studies indicate
that they may be: Smith \& Heckman \shortcite{smi89} showed that about
50\% of nearby powerful FR\,II galaxies show morphological disturbances
from pure elliptical profiles, indicative of galaxy interactions or
mergers, and their optical colours are typically bluer suggesting recent
star formation.

\subsection{Mass dependence of radio and emission line AGN activity} 

Another interesting result of this paper is the fact that radio and
optical AGN appear to trace very different populations of galaxies.  Radio
galaxies are preferentially located in the oldest, most massive galaxies,
whereas emission line AGN are more spread out in mass.

\begin{figure}
\centerline{
\psfig{file=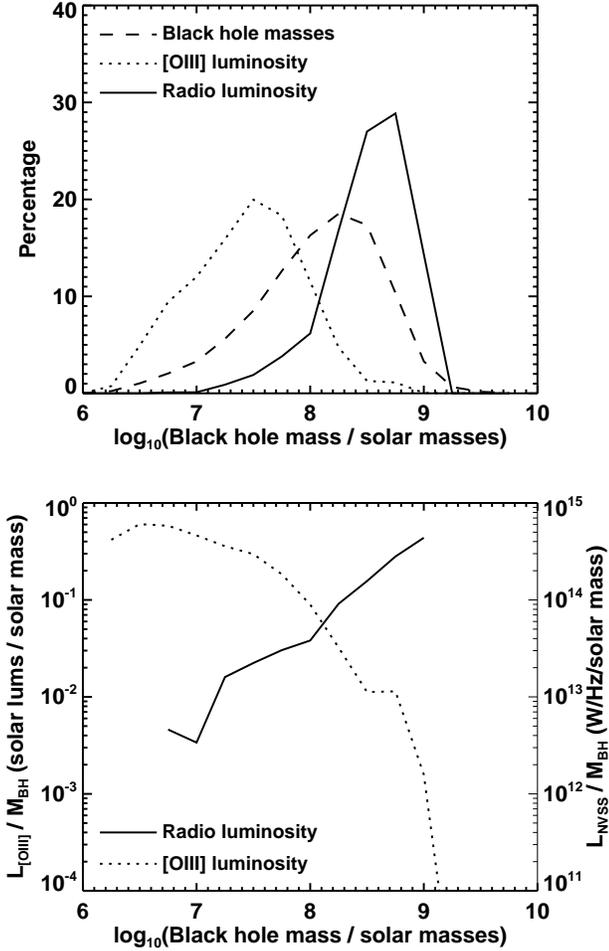,width=8.6cm,clip=}
} 
\caption{\label{lumbhmass} {\it Top:} the distribution, over black hole
mass, of the volume--weighted total mass of all black holes (dashed line),
the volume--weighted total [OIII]~5007 line luminosity from emission--line
selected type-2 AGN (dotted line), and the volume--weighted total radio
luminosity from radio--loud AGN (solid line). {\it Bottom:} the ratio of
the total volume--weighted [OIII] luminosity to the total volume--weighted
mass in black holes (dashed line and left axis), as a function of black
hole mass, compared to the equivalent distribution of total
volume--weighted radio luminosity per total volume--weighted black hole
mass (solid line and right axis).}
\end{figure}

Heckman \etal\ \shortcite{hec04} used the emission--line selected AGN to
investigate the rate at which black holes are currently growing in
galaxies of different masses. They determined how the (volume--weighted)
integrated [OIII]~5007 line luminosity from all emission--line selected
AGN is distributed across black hole mass, and compared this to how the
total black hole mass is distributed. By taking the ratio of these two
distributions, they showed that the average emission line luminosity
output per unit black hole mass is highest at the lowest masses, and falls
steeply towards high masses. The analysis of Heckman \etal\
\shortcite{hec04} is repeated in Figure~\ref{lumbhmass}, demonstrating
essentially the same results (there is a global shift in the normalisation
of the lower panel, compared to Heckman \etal, because of the use of
extinction--corrected [OIII] luminosities in this paper instead of the
directly measured values used by Heckman \etal). Figure~\ref{lumbhmass}
also includes the equivalent plots for the distribution of integrated AGN
radio luminosity across black hole mass (top panel), and the distribution
of radio luminosity per unit black hole mass (lower panel). This clearly
demonstrates that radio and emission--line luminosities are produced by
black holes in very different mass ranges.

By converting emission line luminosities into approximate accretion rates
(see their paper for details), Heckman \etal\ \shortcite{hec04} showed
that at their current accretion rates, black holes of $\sim 10^7
M_{\odot}$ would grow to their current size in about a Hubble time, but
the most massive black holes must have grown substantially faster in the
past. If indeed the emission line luminosity is a good tracer of the
accretion rate, then the low emission line luminosities of the radio--loud
galaxies suggest that their black holes are currently in a very low
accretion rate mode.

A few caveats should be mentioned, however. The observed emission line
luminosity depends upon the ionising output of the AGN and upon the
covering factor of the emission line clouds. Giant elliptical galaxies
have a very low gas content, and so they may have lower emission line
luminosities than would be expected for their level of ionising
radiation. This can explain why the fraction of high luminosity
emission--line AGN falls for the most massive galaxies. An additional
factor is that radio sources may be accreting in a different mode to
emission--line selected AGN (e.g. in advection dominated accretion flows
rather than a standard thin disk). They may thus have larger accretion
rates than optical AGN of the same emission--line luminosity.

It is unlikely, however, that most radio loud AGN are accreting at
substantial rates, because this would imply an unacceptable level of black
hole mass growth in massive galaxies at the present day. The global
mass--doubling time for the black hole population at a given mass can be
written as $t = M_{\rm av} \eta c^2 / L_{\rm av}$, where $M_{\rm av}$ is
the volume--averaged mass density in black holes, $L_{\rm av}$ is the
volume average radio luminosity density, and $\eta$ is an efficiency
factor, given by $L_{\rm rad} = \dot{M} \eta c^2$, for the conversion of
accreted mass energy to radio luminosity. The factor $\eta$ is thus the
product of the efficiency of conversion of accretion mass energy into jet
kinetic energy ($\eta_{\rm acc}$), and the efficiency of conversion of
this into radio luminosity ($\eta_{\rm rad}$); Bicknell \etal\
\shortcite{bic95} estimate that for FR\,I radio sources, the total kinetic
energy output of the radio jet is a factor of 100--1000 higher than the
radio luminosity, and so $\eta_{\rm rad} \sim 10^{-2}$--$10^{-3}$; this
number is also confirmed by the observations of B{\^ i}rzan \etal\
\shortcite{bir04}.  Taking $L_{\rm rad}$ to be $\nu P_{\nu}$ at 1.4\,GHz,
Figure~\ref{lumbhmass} implies that the mass--doubling time associated
with the radio--loud AGN activity is $t \sim 10^{16} \eta$ years for black
hole masses of $\sim 10^9 M_{\odot}$, and even longer for smaller black
holes. The black hole growth (and hence accretion rate) is only
significant if the doubling time is shorter than the Hubble time, ie. if
$\eta \lta 10^{-6}$; even given the result of Bicknell \etal\ this would
still require an accretion efficiency $\eta_{\rm acc} < 10^{-3}$.  Note
also that for a value of $\eta \lta 10^{-6}$ then the most powerful FR\,I
sources, with $L_{\rm 1.4 GHz} \sim 10^{25}$W\,Hz$^{-1}$ would have black
hole growth rates of a few tens of solar masses per year (doubling times
of a few $\times 10^7$ years, corresponding roughly to Eddington limited
accretion). This would require a substantial supply of gas and would be
inconsistent with the hypothesis that the lack of observable emission
lines was entirely due to the low gas content of the host galaxy.

The most likely interpretation is that these are giant elliptical galaxies
with very massive black holes, which formed at early cosmic epoch and
probably contributed to the peak in the quasar population at that
time. Their current activity is associated with a very low accretion rate
(albeit perhaps not as low as would be predicted by their emission line
luminosities alone), which manifests itself in the form of low-power radio
jets but rather little radiation at optical, ultraviolet and infrared
wavelengths. 

The proportion of massive galaxies that are found to be radio--loud AGN
has important implications for the lifecycle of powerful radio
sources. The ages of FR\,II sources can be estimated either by using
synchrotron age--dating of the radio lobes (e.g. Liu, Pooley \& Riley
1992)\nocite{liu92} or directly measured hotspot advance speeds
(e.g. Owsianik, Conway \& Polatidis 1998)\nocite{ows98a}, and these
sources are believed to have lifetimes of typically a few $10^7$ years. At
the radio luminosities of these sources, $L_{\rm 1.4 GHz} \gta
10^{25}$W\,Hz$^{-1}$, the proportion of the highest stellar or black hole
mass galaxies which are radio loud is only a few percent, meaning that
each source would need to be re-triggered only once every one--to--few
Gyrs. This is quite consistent with a relatively high accretion rate
during the active period, which would account for the strong line emission
of FR\,II sources. For the lower radio luminosity FR\,I sources, and even
more so for sources where the entirety of the radio emission arises from
the nucleus, the lifetimes of the radio sources are not known at all. The
fact that $\sim 30$\% of the most massive galaxies are active, however,
implies that however long these lifetimes are, the activity must be
constantly re--triggered so that the galaxy spends over a quarter of its
time in an active state.

In conclusion, emission--line selection of AGN largely picks out a
population which traces the current growth of black holes. This is
dominated by black holes of low and intermediate mass. Radio selection of
AGN (at least at low radio luminosities) picks out the largely dormant
population of the most massive black holes. These are two fundamentally
different modes of black hole activity.

\subsection{Origins of radio--loud AGN activity: gas cooling?} 

Burns \shortcite{bur90} showed that central massive galaxies in clusters
or groups of galaxies with a cooling flow are much more likely to be
radio--loud AGN than galaxies of similar mass not centred on a cooling
flow, suggesting that concentrated cooling atmospheres can stimulate
radio--loud AGN activity. Elliptical galaxies are surrounded by
atmospheres of hot gas (see the review by Mathews \& Brighenti 2003, and
references therein) \nocite{mat03b} and it is intriguing that the rate at
which gas cools out of these atmospheres has almost exactly the same
dependence on black hole mass as has the radio--loud AGN fraction.

Consider a small region of gas, of proton density $n$, volume $\delta V$
and temperature $T$. The mass cooling rate of this gas is proportional to
the mass of the gas ($\delta M_{\rm gas} = n m_p \delta V$, where $m_p$ is
the proton mass), divided by the cooling time. The cooling time is given
by the total energy of the gas ($3nkT\delta V$) divided by the rate at
which energy is radiated away; since much of the cooling occurs at X--ray
wavelengths, the energy radiation rate is roughly proportional to the
X--ray luminosity of that region ($\delta L_X$). Therefore, the mass
cooling rate of gas within volume $\delta V$ is given by:

\begin{displaymath}
\delta \dot{M}_{\rm cool} \propto \frac{\delta M_{\rm gas}}{t_{\rm cool}} 
\propto \frac{n m_p \delta V}{3nkT\delta V/\delta L_X} \propto
\frac{\delta L_X}{T} 
\end{displaymath}

Integrating over the entire envelope of the elliptical, if the atmosphere
is isothermal then the total mass cooling rate is given by $\dot{M}_{\rm
cool} \propto L_X / T$, where $L_X$ is the total X--ray luminosity of the
elliptical.  For an isothermal gas then $T \propto \sigma^2$, where
$\sigma$ is the velocity dispersion. Also, for luminous ellipticals (which
represent those in which radio sources are found) the X--ray luminosity is
observed to vary as the square of the optical luminosity, $L_X \propto
L_{\rm opt}^2$ \cite{osu01,mat03b}. The Faber--Jackson relation dictates
that the optical luminosity goes as $L_{\rm opt} \propto \sigma^4$, and
therefore $L_X \propto \sigma^8$. Note that Mahdavi \& Geller
\shortcite{mah01} have confirmed that elliptical galaxies do indeed show
such a tight correlation between their X--ray luminosity and velocity
dispersion (the exponent that they derive observationally from a
57--galaxy sample is $10^{+4}_{-2}$).

These relations, together with the black hole mass versus velocity
dispersion relation, $M_{\rm BH} \propto \sigma^4$, give the following:

\begin{displaymath}
\dot{M}_{\rm cool} \propto L_X / T \propto \sigma^6 \propto M_{\rm
BH}^{1.5}
\end{displaymath}

This is within the range of the $1.6 \pm 0.1$ exponent derived for the
radio--loud AGN fraction dependence on black hole mass, and suggests that
for these low power radio sources the fueling of the AGN activity may be
related to the cooling for gas out of the hot atmospheres of the host
galaxies. In terms of numbers, for a $10^{11} M_{\odot}$ elliptical the
actual cooling rate implied by the above calculation is of order a solar
mass per year (although observations suggest that the actual cooling rates
are lower; see Mathews \& Brighenti 2003 and references therein). For the
radiative accretion efficiency of $\eta \sim 10^{-3}$, corresponding to
the ratio between radio luminosity and total jet kinetic energy suggested
by Bicknell \etal\ \shortcite{bic95}, only $10^{-3} M_{\odot}$yr$^{-1}$ of
gas is required to fuel the AGN, so cooling could easily provide enough
gas. 

\subsection{The role of AGN feedback}

There has been much interest recently in the question of whether the
heating effect of AGN activity can balance the cooling of the gas, and
thus whether AGN activity, particularly radio--loud AGN activity, can play
an important feedback role. This has largely been motivated by the
observed effects of radio--loud AGN on their environments at galaxy
cluster scales (e.g. Fabian \etal\ 2003)\nocite{fab03}, but may be equally
important in the haloes of individual galaxies (cf. Mathews \& Brighenti
2003)\nocite{mat03b}.

The rate at which energy is radiated from the hot atmosphere is
proportional to the X--ray luminosity, which as shown above goes as $L_X
\propto \sigma^8 \propto M_{\rm BH}^2$. If feedback is to work then the
time--averaged rate at which the AGN inputs energy back into its
environment must balance that. This depends upon the fraction of time for
which the AGN is active multiplied by the energy output of the AGN when it
is active. If it is assumed that all galaxies of given black hole mass go
through a radio-loud AGN phase, then the fraction of time for which an AGN
is on is simply given by $f_{\rm radio-loud}$, which increases as $M_{\rm
BH}^{1.6}$. In this paper it has been shown that the radio luminosity of a
radio--loud AGN is essentially independent of black hole mass. Thus, if
radio luminosity were a direct measure of the energy output of an AGN, the
time--averaged heating rate of AGN would go as $M_{\rm BH}^{1.6}$.

In fact, radio luminosity is not a direct measure of the kinetic energy
input into the radio jets, since numerous other environmental factors play
a role. A robust calculation of the energy input rate of the AGN is not
possible, due to a lack of knowledge of the lifetimes and radiative
efficiencies of FR\,I type sources. However, using the FR\,II sources as a
basis, the argument below demonstrates that the time--averaged heating may
well go as a higher power of black hole mass, and thus that a balance
between AGN heating and gas cooling in these radio--loud AGN is feasible.

For FR\,II radio sources, models of the growth of the radio sources
(e.g. Kaiser \etal\ 1997a,b; Snellen \etal\
2000)\nocite{kai97a,kai97b,sne00} show that for a fixed jet kinetic energy
input, once the source has expanded beyond the central core of the galaxy
($\sim 1$\,kpc; this happens on a timescale of $\sim 10^5$ years) then the
radio luminosity falls as the source expands into a progressively lower
density environment. For most reasonable assumptions, the radio luminosity
is expected to fall roughly as the size of the source ($D$) to a power
between $-0.4$ and $-0.5$. For a radial density distribution of $\rho
\propto r^{-\beta}$, Kaiser \etal\ \shortcite{kai97a} show that the size
of the source grows with time as $D \propto t^{3/(5-\beta)}$, and argue
that $\beta$ has a value of a little below 2. Therefore, the radio
luminosity of FR\,II sources falls off with age roughly as $t^{-0.4}$, so
for a given radio source the observed radio luminosity needs to be
multiplied up by a factor of age$^{0.4}$ in order to derive the intrinsic
jet power, and hence the energy output of the AGN.  The fraction of
radio--loud AGN is higher in galaxies with more massive central black
holes, and (again assuming that all galaxies of given black hole mass go
through a radio--loud AGN phase) this could either be because the AGN in
these galaxies are re-triggered more often, or because when they are
triggered they live longer; the current data cannot distinguish between
these two possibilities. If this is entirely due to longer lifetimes, then
the average age of the radio sources observed will be proportional to
their lifetime, which in turn will be proportional to the fraction of time
that a black hole of that mass is active, ie age $\propto M_{\rm
BH}^{1.6}$. Therefore, for sources of a given radio luminosity, the energy
output of the AGN will scale as age$^{0.4} \propto M_{\rm
BH}^{0.6}$. Multiplying this by the fraction of time for which the AGN is
active, the energy output scales as $M_{\rm BH}^{2.2}$. This is comparable
to (or greater than) that required to counterbalance the gas cooling.

Many aspects of this calculation are uncertain; in particular, if the
higher fraction of radio--loud AGN in more massive galaxies is due to more
frequent re-triggering rather than longer lifetimes then the extra $M_{\rm
BH}^{0.6}$ factor is lost. Also, it is not known whether the luminosity
evolution of FR\,I sources with time follows that of the FR\,IIs, although
for the extended sources a similar principal of decreasing luminosity as
the source expands into lower densities is expected to apply. However this
calculation does demonstrate that a balance between AGN heating and gas
cooling in these radio--loud AGN is certainly possible.

A final intriguing observation is that if the radio--loud AGN fraction
versus stellar mass and black hole mass relations are extrapolated upwards
(noting that the flattening off at high masses in the latter relation is
likely to be due to errors in the black hole mass estimates), then at a
stellar mass of $10^{12} M_{\odot}$ or a black hole mass of $\sim 3 \times
10^9 M_{\odot}$ essentially 100\% of galaxies would be radio loud. This
means that any galaxy of such a mass must be radio--emitting at all
times. Is it a co-incidence that these masses correspond roughly to the
most massive galaxy and the most massive black hole that exist in the
nearby Universe? Could the very presence of this continual radio source
activity be the reason why galaxies do not grow any larger?

\section*{Acknowledgements} 

PNB would like to thank the Royal Society for generous financial support
through its University Research Fellowship scheme. JB acknowledges receipt
of an ESA post-doctoral fellowship. The research makes use of the SDSS
Archive, funding for the creation and distribution of which was provided
by the Alfred P. Sloan Foundation, the Participating Institutions, the
National Aeronautics and Space Administration, the National Science
Foundation, the U.S. Department of Energy, the Japanese Monbukagakusho,
and the Max Planck Society.  The research uses the NVSS and FIRST radio
surveys, carried out using the National Radio Astronomy Observatory Very
Large Array: NRAO is operated by Associated Universities Inc., under
co-operative agreement with the National Science Foundation. The authors
thank the referee, Chris Simpson, for a number of helpful suggestions.
\vspace*{-0.2cm}

\bibliography{pnb} 

\begin{thebibliography}{}

\bibitem[\protect\citename{Antonucci{\ }}{1993}]{ant93}
Antonucci~R.,  1993, ARA\&A, 31, 473

\bibitem[\protect\citename{Archibald et~al.{\ }}{2002}]{arc02a}
Archibald~E.~N.,  Dunlop~J.~S.,  Jimenez~R.,  Fraica~A. C.~S.,  McLure~R.~J.,
   Hughes~D.,  2002, MNRAS, 336, 353

\bibitem[\protect\citename{Auriemma et~al.{\ }}{1977}]{aur77}
Auriemma~C.~G.,  Perola~G.,  Ekers~R.,  Fanti~R.,  Lari~C.,  Jaffe~W.,
  Ulrich~M.,  1977, A\&A, 57, 41

\bibitem[\protect\citename{{B{\^ i}rzan} et~al.{\ }}{2004}]{bir04}
{B{\^ i}rzan}~L.,  {Rafferty}~D.~A.,  {McNamara}~B.~R.,  {Wise}~M.~W.,
  {Nulsen}~P.~E.~J.,  2004, ApJ, 607, 800

\bibitem[\protect\citename{Baldwin et~al.{\ }}{1981}]{bal81}
Baldwin~J.~A.,  Phillips~M.~M.,    Terlevich~R.,  1981, PASP, 93, 5

\bibitem[\protect\citename{Barthel \& Arnaud{\ }}{1996}]{bar96a}
Barthel~P.~D.,  Arnaud~K.~A.,  1996, MNRAS, 283, L45

\bibitem[\protect\citename{Baum et~al.{\ }}{1995}]{bau95}
Baum~S.~A.,  Zirbel~E.~L.,    {O'Dea}~C.~P.,  1995, ApJ, 451, 88

\bibitem[\protect\citename{Becker et~al.{\ }}{1995}]{bec95}
Becker~R.~H.,  White~R.~L.,    Helfand~D.~J.,  1995, ApJ, 450, 559

\bibitem[\protect\citename{Best{\ }}{2004}]{bes04a}
Best~P.~N.,  2004, MNRAS, 351, 70

\bibitem[\protect\citename{Best et~al.{\ }}{2005}]{bes05a}
Best~P.~N.,  Kauffmann~G.,  Heckman~T.~M.,  Ivezi{\'c}   \v{Z}.  2005, MNRAS,
  submitted

\bibitem[\protect\citename{Bicknell{\ }}{1995}]{bic95}
Bicknell~G.~V.,  1995, ApJ Supp., 101, 29

\bibitem[\protect\citename{B{\"o}hringer et~al.{\ }}{1993}]{boh93}
B{\"o}hringer~H.,  Voges~W.,  Fabian~A.~C.,  Edge~A.~C.,    Neumann~D.~M.,
  1993, MNRAS, 264, L25

\bibitem[\protect\citename{Brinchmann et~al.{\ }}{2004}]{bri04b}
Brinchmann~J.,  Charlot~S.,  Heckman~T.,  Kauffmann~G.,  Tremonti~C.,
  White~S. D.~M.,  2004, astro-ph/0406220

\bibitem[\protect\citename{{Burns}{\ }}{1990}]{bur90}
{Burns}~J.~O.,  1990, AJ, 99, 14

\bibitem[\protect\citename{{Cao} \& {Rawlings}{\ }}{2004}]{cao04}
{Cao}~X.,  {Rawlings}~S.,  2004, MNRAS, 349, 1419

\bibitem[\protect\citename{{Colless et~al.}{\ }}{2001}]{col01}
{Colless~M.~M. et~al.} 2001, MNRAS, 328, 1039

\bibitem[\protect\citename{Condon et~al.{\ }}{1998}]{con98}
Condon~J.~J.,  Cotton~W.~D.,  Greisen~E.~W.,  Yin~Q.~F.,  Perley~R.~A.,
  Taylor~G.~B.,    Broderick~J.~J.,  1998, AJ, 115, 1693

\bibitem[\protect\citename{{De Bruyn} \& Wilson{\ }}{1978}]{deb78}
{De Bruyn}~A.~G.,  Wilson~A.~S.,  1978, A\&A, 64, 433

\bibitem[\protect\citename{Fabian{\ }}{1999}]{fab99}
Fabian~A.~C.,  1999, MNRAS, 308, L39

\bibitem[\protect\citename{Fabian et~al.{\ }}{2003}]{fab03}
Fabian~A.~C.,  Sanders~J.~S.,  Allen~S.~W.,  Crawford~C.~S.,  Iwasawa~K.,
  Johnstone~R.~M.,  Schmidt~R.~W.,    Taylor~G.~B.,  2003, MNRAS, 344, L43

\bibitem[\protect\citename{Fanaroff \& Riley{\ }}{1974}]{fan74}
Fanaroff~B.~L.,  Riley~J.~M.,  1974, MNRAS, 167, 31P

\bibitem[\protect\citename{Ferrarese \& Ford{\ }}{2004}]{fer04}
Ferrarese~L.,  Ford~H.,  2004, astro-ph/0411247

\bibitem[\protect\citename{Ferrarese \& Merritt{\ }}{2000}]{fer00}
Ferrarese~L.,  Merritt~D.,  2000, ApJ, 539, L9

\bibitem[\protect\citename{{Gebhardt et~al.}{\ }}{2000}]{geb00}
{Gebhardt~K. et~al.} 2000, ApJ, 539, L13

\bibitem[\protect\citename{{H{\" a}ring} \& {Rix}{\ }}{2004}]{har04}
{H{\" a}ring}~N.,  {Rix}~H.-W.,  2004, ApJ, 604, L89

\bibitem[\protect\citename{{Hao et~al.}{\ }}{2005}]{hao05}
{Hao~L. et~al.} 2005, AJ, in press; astro-ph/0501059

\bibitem[\protect\citename{Hartwick \& Schade{\ }}{1990}]{har90}
Hartwick~F. D.~A.,  Schade~D.,  1990, ARA\&A, 28, 437

\bibitem[\protect\citename{Heckman et~al.{\ }}{2004}]{hec04}
Heckman~T.~M.,  Kauffmann~G.,  Brinchmann~J.,  Charlot~S.,  Tremonti~C.,
  White~S.~D.,  2004, ApJ, 613, 109

\bibitem[\protect\citename{Hine \& Longair{\ }}{1979}]{hin79}
Hine~R.~G.,  Longair~M.~S.,  1979, MNRAS, 188, 111

\bibitem[\protect\citename{Kaiser \& Alexander{\ }}{1997}]{kai97a}
Kaiser~C.~R.,  Alexander~P.,  1997, MNRAS, 286, 215

\bibitem[\protect\citename{Kaiser et~al.{\ }}{1997}]{kai97b}
Kaiser~C.~R.,  Dennett-Thorpe~J.,    Alexander~P.,  1997, MNRAS, 292, 723

\bibitem[\protect\citename{Kauffmann et~al.{\ }}{2004}]{kau04}
Kauffmann~G.,  White~S. D.~M.,  Heckman~T.~M.,  M{\'e}nard~B.,  Brinchmann~J.,
  Charlot~S.,  Tremonti~C.,    Brinkmann~J.,  2004, MNRAS, 353, 713

\bibitem[\protect\citename{{Kauffmann et~al.}{\ }}{2003a}]{kau03c}
{Kauffmann~G. et~al.} 2003a, MNRAS, 346, 1055

\bibitem[\protect\citename{{Kauffmann et~al.}{\ }}{2003b}]{kau03a}
{Kauffmann~G. et~al.} 2003b, MNRAS, 341, 33

\bibitem[\protect\citename{{Kauffmann et~al.}{\ }}{2003c}]{kau03b}
{Kauffmann~G. et~al.} 2003c, MNRAS, 341, 54

\bibitem[\protect\citename{King{\ }}{2003}]{kin03}
King~A.,  2003, ApJ, 596, L27

\bibitem[\protect\citename{Kormendy \& Richstone{\ }}{1995}]{kor95}
Kormendy~J.,  Richstone~D.,  1995, ARA\&A, 33, 581

\bibitem[\protect\citename{LaValley et~al.{\ }}{1992}]{lav92}
LaValley~M.,  Isobe~T.,    Feigelson~E.,  1992, BAAS, 24, 839

\bibitem[\protect\citename{Ledlow \& Owen{\ }}{1996}]{led96}
Ledlow~M.~J.,  Owen~F.~N.,  1996, AJ, 112, 9

\bibitem[\protect\citename{Liu et~al.{\ }}{1992}]{liu92}
Liu~R.,  Pooley~G.,    Riley~J.~M.,  1992, MNRAS, 257, 545

\bibitem[\protect\citename{Magorrian et~al.{\ }}{1998}]{mag98a}
Magorrian~J.,  Tremaine~S.,  Richstone~D.,  Bender~R.,  Bower~G.,  Dressler~A.,
   Faber~S.,  Gebhardt~K.,  Green~R.,  Grillmair~C.,  Kormendy~J.,    Lauer~T.,
   1998, AJ, 115, 2285

\bibitem[\protect\citename{Mahdavi \& Geller{\ }}{2001}]{mah01}
Mahdavi~A.,  Geller~M.~J.,  2001, ApJ, 554, L129

\bibitem[\protect\citename{{Marconi} \& {Hunt}{\ }}{2003}]{mar03}
{Marconi}~A.,  {Hunt}~L.~K.,  2003, ApJ, 589, L21

\bibitem[\protect\citename{Mathews \& Brighenti{\ }}{2003}]{mat03b}
Mathews~W.~G.,  Brighenti~F.,  2003, ARA\&A, 41, 191

\bibitem[\protect\citename{Matthews et~al.{\ }}{1964}]{mat64}
Matthews~T.~A.,  Morgan~W.~W.,    Schmidt~M.,  1964, ApJ, 140, 35

\bibitem[\protect\citename{McCarthy{\ }}{1993}]{mcc93}
McCarthy~P.~J.,  1993, ARA\&A, 31, 639

\bibitem[\protect\citename{McGaugh \& {de Blok}{\ }}{1997}]{mcg97}
McGaugh~S.~S.,  {de Blok}~W. J.~C.,  1997, ApJ, 481, 689

\bibitem[\protect\citename{Miller et~al.{\ }}{2003}]{mil03a}
Miller~C.~J.,  Nichol~R.~C.,  Gom{\'e}z~P.~L.,  Hopkins~A.~M.,    Bernardi~M.,
  2003, ApJ, 597, 142

\bibitem[\protect\citename{{O'Sullivan} et~al.{\ }}{2001}]{osu01}
{O'Sullivan}~E.,  Forbes~D.~A.,    Ponman~T.~J.,  2001, MNRAS, 328, 461

\bibitem[\protect\citename{Owsianik et~al.{\ }}{1998}]{ows98a}
Owsianik~I.,  Conway~J.~E.,    Polatidis~A.~G.,  1998, A\&A, 336, L370

\bibitem[\protect\citename{Rawlings \& Saunders{\ }}{1991}]{raw91b}
Rawlings~S.,  Saunders~R.,  1991, Nat, 349, 138

\bibitem[\protect\citename{Rawlings et~al.{\ }}{1989}]{raw89}
Rawlings~S.,  Saunders~R.,  Eales~S.~A.,    Mackay~C.~D.,  1989, MNRAS, 240,
  701

\bibitem[\protect\citename{{Sadler} et~al.{\ }}{1989}]{sad89}
{Sadler}~E.~M.,  {Jenkins}~C.~R.,    {Kotanyi}~C.~G.,  1989, MNRAS, 240, 591

\bibitem[\protect\citename{{Sadler et~al.}{\ }}{2002}]{sad02}
{Sadler~E.~M. et~al.} 2002, MNRAS, 329, 227

\bibitem[\protect\citename{Sanders et~al.{\ }}{1988}]{san88}
Sanders~D.~B.,  Soifer~B.~T.,  Elias~J.~H.,  Neugebauer~G.,    Matthews~K.,
  1988, ApJ, 328, L35

\bibitem[\protect\citename{Silk \& Rees{\ }}{1998}]{sil98}
Silk~J.,  Rees~M.~J.,  1998, A\&A, 331, L1

\bibitem[\protect\citename{Simpson{\ }}{2005}]{sim05}
Simpson~C.,  2005, MNRAS, in press; astro-ph/0503500

\bibitem[\protect\citename{Smith \& Heckman{\ }}{1989}]{smi89}
Smith~E.~P.,  Heckman~T.~M.,  1989, ApJ, 341, 658

\bibitem[\protect\citename{Snellen et~al.{\ }}{2000}]{sne00}
Snellen~I. A.~G.,  Schilizzi~R.~T.,  Miley~G.~K.,  {de Bruyn}~A.~G.,
  Bremer~M.~N.,    R{\"o}ttgering~H. J.~A.,  2000, MNRAS, 319, 445

\bibitem[\protect\citename{{Stoughton et~al.}{\ }}{2002}]{sto02}
{Stoughton~C. et~al.} 2002, AJ, 123, 485

\bibitem[\protect\citename{{Tremaine et~al.}{\ }}{2002}]{tre02}
{Tremaine~S. et~al.} 2002, ApJ, 574, 740

\bibitem[\protect\citename{Tremonti et~al.{\ }}{2004}]{tre04}
Tremonti~C.~A.,  Heckman~T.~M.,  Kauffmann~G.,  Brinchmann~J.,  Charlot~S.,
  White~S. D.~M.,  Seibert~M.,  Peng~E.~W.,  Schlegel~D.~J.,  Uomoto~A.,
  Fukugita~M.,    Brinkmann~J.,  2004, ApJ, 613, 898

\bibitem[\protect\citename{Wills et~al.{\ }}{2004}]{wil04}
Wills~K.~A.,  Morganti~R.,  Tadhunter~C.~N.,  Robinson~T.~G.,
  {Villar-Mart{\'\i}n}~M.,  2004, MNRAS, 347, 771

\bibitem[\protect\citename{Xu et~al.{\ }}{1999}]{xu99}
Xu~C.,  Livio~M.,    Baum~S.,  1999, AJ, 118, 1169

\bibitem[\protect\citename{{York et~al.}{\ }}{2000}]{yor00}
{York~D.~G. et~al.} 2000, AJ, 120, 1579

\bibitem[\protect\citename{Zirbel \& Baum{\ }}{1995}]{zir95}
Zirbel~E.~L.,  Baum~S.~A.,  1995, ApJ, 448, 521

\end{thebibliography}
\bibliographystyle{mn} 

\label{lastpage}

\end{document}